\documentclass[aps,prx,twocolumn,superscriptaddress]{revtex4-1}

\usepackage{amssymb}
\usepackage{amsmath}
\usepackage{graphicx}
\usepackage[caption=false]{subfig}
\usepackage{floatrow}
\usepackage[dvipsnames]{xcolor}
\usepackage[version=3]{mhchem}
\usepackage{footnote}
\usepackage{bm}
\usepackage{dsfont}
\usepackage{xr-hyper}
\usepackage[hidelinks]{hyperref}
\usepackage{cleveref}
\usepackage{soul}
\usepackage[normalem]{ulem}

\usepackage{booktabs}
\crefname{equation}{Eq.}{Eqs.}
\Crefname{equation}{Equation}{Equations}
\crefname{figure}{Fig.}{Figs.} 
\Crefname{figure}{Figure}{Figures}
\crefname{section}{Sect.}{Sects.}
\Crefname{section}{Section}{Sections}
\crefname{table}{Table}{Tables}
\crefname{appsec}{}{Appendices}
\graphicspath{{Figures/}}

\newcommand{\abs}[1]{\ensuremath{\left|#1\right|}}
\newcommand{\ket}[1]{\ensuremath{\left|#1\right\rangle}}

\newcommand{\braket}[2]{\ensuremath{\left\langle#1|#2\right\rangle}}
\newcommand{\ketbra}[2]{\ensuremath{\left|#1\right\rangle\left\langle#2\right|}}
\newcommand{\comm}[2]{\ensuremath{\left[#1,#2\right]}}
\newcommand{\acomm}[2]{\ensuremath{\left\lbrace#1,#2\right\rbrace}}
\newcommand{\matrixel}[3]{\ensuremath{\left\langle#1|#2|#3\right\rangle}}

\definecolor{catcolor}{rgb}{0.,0.5,0.5}

\usepackage{tabularx}

\begin{document}

\floatsetup[figure]{style=plain,subcapbesideposition=top}

\title{Superconducting coupler with exponentially large on-off ratio}

\author{Catherine Leroux}
\email{Catherine.Leroux@USherbrooke.ca}
\affiliation{Institut quantique \& D\'epartement de Physique, Universit\'e de Sherbrooke, Sherbrooke J1K 2R1 QC, Canada}

\author{Agustin Di Paolo}\thanks{adipaolo@mit.edu} 
\affiliation{Institut quantique \& D\'epartement de Physique, Universit\'e de Sherbrooke, Sherbrooke J1K 2R1 QC, Canada}

\author{Alexandre Blais}
\affiliation{Institut quantique \& D\'epartement de Physique, Universit\'e de Sherbrooke, Sherbrooke J1K 2R1 QC, Canada}
\affiliation{Canadian Institute for Advanced Research, Toronto, ON, Canada}

\date{\today}

\begin{abstract}
    Tunable two-qubit couplers offer an avenue to mitigate errors in multiqubit superconducting quantum processors. However, most couplers operate in a narrow frequency band and target specific couplings, such as the spurious~$ZZ$ interaction. We introduce a superconducting coupler that alleviates these limitations by suppressing all two-qubit interactions with an exponentially large on-off ratio and without the need for fine-tuning. Our approach is based on a bus mode supplemented by an ancillary nonlinear resonator mode. Driving the ancillary mode leads to a coupler-state-dependent field displacement in the resonator which, in turn, results in an exponential suppression of real and virtual two-qubit interactions with respect to the drive power. A superconducting circuit implementation supporting the proposed mechanism is presented.  
\end{abstract}
                 
\maketitle

\section{Introduction} 

Two-qubit couplers are useful components for quantum information processing as they enable fast and high-fidelity operations between qubits while reducing crosstalk during idle times. Several superconducting coupler designs have been theoretically proposed and experimentally implemented~\cite{Tsai2007,Schoelkopf2007,Devoret2017,Oliver2018,Poletto2021,Finck2021,Marcus2019,Homann2013,Martinis2014,arute2019quantum,Gambetta2016,Schmidt2017,Li2018,Steele2018,Yan2021,Eichler2020,Oliver2021,Poletto2021_2}. These devices can, in principle, offer precise control of two-qubit interactions while helping to mitigate frequency crowding effects in multiqubit processors such as to improve gate speed and fidelity. There are, however, limitations to the performance of current couplers. For instance, while couplers are designed to activate interactions between qubits on-demand, spurious interactions can remain active when the coupler is tuned to its `off' state. A common example is the ubiquitous always-on cross-Kerr, or~$ZZ$, coupling~\cite{Wilhelm2011,Olivier2018,Houck2018,Devoret2017,Plourde2020,Yu2020}. A second difficulty is that the coupler's on-off ratio is often sensitive to first order in a control parameter, such as an external magnetic flux, thus requiring fine-tuning and frequent calibration. Couplers which do not rely on frequency tuning do not suffer from this, but the lack of tunability comes with its own set of challenges such as large crosstalk errors during idle times.
Finally, the impact of these effects could be exacerbated in multiqubit devices where frequency shifts from spectator qubits can counteract fine-tuning. 

\begin{figure}[b]
    \centering
    \includegraphics[width=0.85\textwidth]{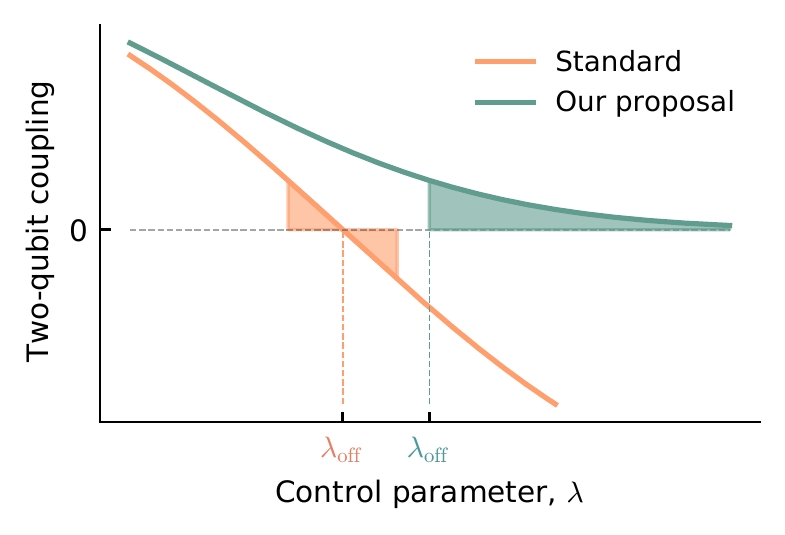}
    \caption{Two-qubit coupling strength vs control parameter. For standard coupler designs, the on-off ratio depends linearly with respect to a control parameter~$\lambda$ (orange line). This results in linear sensitivity to noise in the control parameter. Our coupler design implements a two-qubit coupling that can be exponentially suppressed with respect to the control parameter (green line). Therefore, sensitivity to noise in the `off' state of the coupler is exponentially small, and the need for fine-tuning is relaxed.
    }
    \label{fig:proposal}
\end{figure}

Here, we alleviate these issues by introducing a tunable coupler with an exponentially large on-off ratio and that does not require fine-tuning of the coupler or qubit parameters. This is realized by adapting some of the ideas of protected qubits to coupler designs, as illustrated in~\cref{fig:proposal}. Broadly speaking, the large on-off ratio is achieved by connecting a bus mode that wires the two qubits to an ancillary driven nonlinear resonator (NLR), in such a way that the bus transition matrix elements that control two-qubit interactions vanish exponentially with respect to the amplitude of the drive on the ancillary system. This key feature renders the coupler, which includes the bus and the NLR modes, exponentially insensitive to noise and relaxes the need for fine-tuning. 

This paper is organized as follows. In~\cref{sec:model} we describe the physical mechanism enabling the exponential suppression of two-qubit interactions and introduce a model Hamiltonian realizing this mechanism. In~\cref{sec:numerics}, we then report numerical results demonstrating the exponential suppression of qubit-qubit interactions mediated by the coupler and discuss implications in the context of large-scale processors as well as some of the limitations of the proposed design. Finally, we discuss a superconducting circuit implementation in~\cref{sec:implementation}. 

\begin{figure*}[t]
    \centering
    \includegraphics[width=0.9\textwidth]{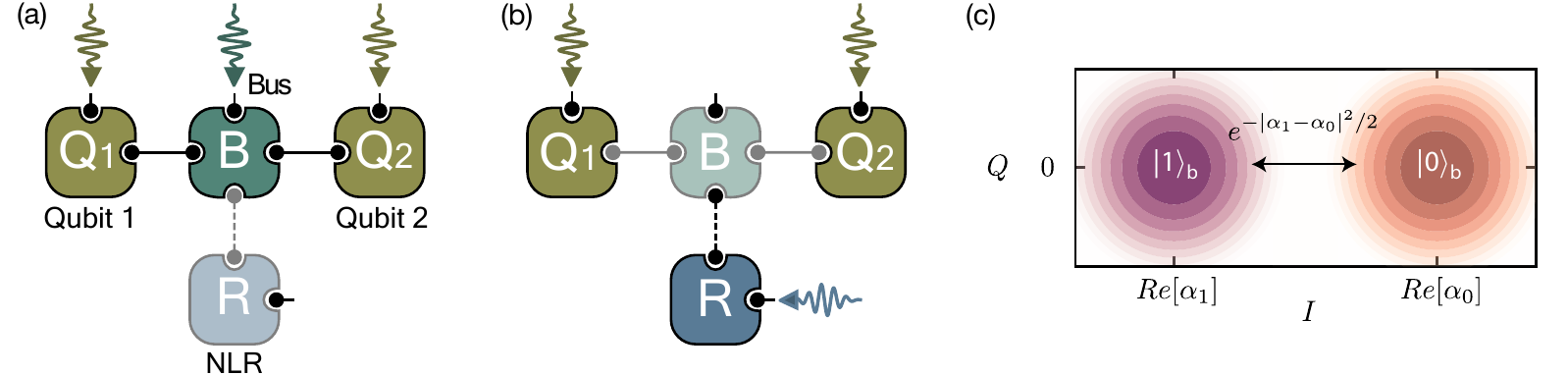}
    \caption{Illustration of the proposed superconducting coupler. 
    a) In the `on' state of the coupler the NLR does not participate in the two-qubit interactions mediated by the bus mode. Local drives on the qubits or bus activate two-qubit gates. b) In the `off' state, the NLR is subject to a microwave drive which strongly suppresses two-qubit interactions that are mediated by the bus. c) Metapotential of the NLR for the bus states~$\ket{0}_b$ (orange) and~$\ket{1}_b$ (purple). Here~$\delta/2\pi = - 5.0$ MHz,~$\chi/2\pi = -20.0$ MHz and~$K_r = 0$. To help visualization, the metapotential~$\mathcal{E}(I,Q)$ is renormalized as~$\mathcal{E} \cdot \left[(\delta+n\chi)\abs{\alpha_0/4}^2\right]^{-1}$ for the~$n$th bus state and white corresponds to unity. The global minima of the metapotentials for~$n\geq 2$ are close to the global minimum of the~$n=1$ metapotential but not shown for simplicity. 
    }
    \label{fig:decoupling_mechanism}
\end{figure*}

\section{Working principle and Hamiltonian model \label{sec:model}}

\Cref{fig:decoupling_mechanism} a-b) schematically illustrates the proposed device consisting of two qubits, labeled~$Q_1$ and~$Q_2$ coupled by a bus mode~$B$ also connected to a NLR mode~$R$. In the absence of a drive on the NLR, the system reduces to a standard circuit QED setup where the bus mediates energy-exchange interactions between the qubits~\cite{Schoelkopf2004,Blais2020,Olivier2018,Houck2018,Devoret2017,Plourde2020,Yu2020}. We assume that the qubit-bus interactions (full lines) can be modeled by a Jaynes-Cummings-type Hamiltonian. Instead, the bus-NLR interaction (dashed line) is engineered such that, upon driving the second mode, the resonator field undergoes a bus-state-dependent displacement characteristic of a longitudinal interaction. As a result, a distinct resonator coherent states~$\ket{\alpha_n}_r$ is associated with each bus eigenstates~$\ket{n}_b$, such that the states~$\ket{\psi_n} = \ket{n}_b\ket{\alpha_n}_r$ are stabilized. Then, transitions between the~$m$th and~$n$th low-energy eigenstates of the bus are suppressed in the coherent-state amplitude 
\begin{align}
    \ketbra{n}{m}_b \otimes 1_r \xrightarrow[]{} e^{-\abs{\alpha_n-\alpha_m}^2/2}\ \ketbra{\psi_n}{\psi_m}. 
    \label{eq:nm_transition}
\end{align}
Because all two-qubit interactions are mediated by real or virtual transitions amongst the bus eigenstates, suppressing these transitions robustly switches off all interactions mediated by the coupler. As discussed below, if the bus mode is constrained to its ground state, only the virtual transitions of the form~$0\rightarrow n$ need to be suppressed for all~$n$. We also note that this mechanism is reminiscent of the strategy used to protect cat qubits from spurious bit flips~\cite{Munro1999, Blais2017, Mirrahimi2014}.

An effective Hamiltonian realizing this decoupling mechanism can be put in the form 
\begin{align}
    \hat{H} = \sum_{j=1}^2 \hat{H}_{j} + \hat{H}_b + \hat{H}_r + \sum_{j=1}^2\hat{H}_{jb} + \hat{H}_{br} + \hat{H}_{bnlr}, 
    \label{eq:Hlab}
\end{align}
where
\begin{align}
    & \hat{H}_{j}/\hbar = \omega_{j}  \hat{q}_j^{\dagger}\hat{q}_j + \frac{K_j}{2}\hat{q}_j^{\dagger 2}\hat{q}_j^2,  \\
    & \hat{H}_b/\hbar = \omega_b \hat{b}^{\dagger}\hat{b} + \frac{K_b}{2}\hat{b}^{\dagger 2}\hat{b}^2,
\end{align}
are the qubits ($j=1,2$) and bus Hamiltonians modeled 
as Kerr-nonlinear oscillators, and
\begin{align}
    &  \hat{H}_r/\hbar = \omega_r \hat{r}^{\dagger}\hat{r}  - \varepsilon(t) e^{-i\omega_d t}\hat{r}^{\dagger} - \varepsilon^*(t) e^{i\omega_d t}\hat{r},
\end{align}
is the NLR Hamiltonian subject to a drive of amplitude~$\varepsilon(t)$ and frequency~$\omega_d$. In these expressions,~$\hat{q}_1$,~$\hat{q}_2$,~$\hat{b}$,~$\hat{r}$ are the annihilation operators of~$Q_1$,~$Q_2$,~$B$, and~$R$ with mode frequencies~$\omega_{1}$,~$\omega_{2}$,~$\omega_{b}$,~$\omega_{r}$, and anharmonicities~$K_{1}$,~$K_{2}$,~$K_b$,  respectively. Although our Hamiltonian model is formulated for the case of transmon qubits~\cite{Koch2007}, which can be described as Kerr nonlinear oscillators, our coupling scheme is in principle applicable to other qubit modalities. The qubits interact with the bus mode through a Jaynes-Cummings-type Hamiltonian of the form
\begin{align}
    & \hat{H}_{jb}/\hbar = g_j   \left(\hat{q}_j^{\dagger}\hat{b} +  \hat{b}^{\dagger}\hat{q}_j\right),  
\end{align}
where~$g_{j}$ is the coupling strength, while the bus mode interacts with the NLR via the cross-Kerr coupling Hamiltonian
\begin{align}
    & \hat{H}_{br}/\hbar = \chi \hat{b}^{\dagger}\hat{b}\hat{r}^{\dagger}\hat{r}
    \label{eq:Hlab2}
\end{align}
which makes the NLR's frequency conditional on the bus state via the dispersive shift~$\chi$. For the last term of~\cref{eq:Hlab}, we assume the form
\begin{align}
    & \hat{H}_{bnlr}/\hbar = \sum_n \frac{K_{r}}{2} \ketbra{n}{n}_b \otimes \hat{r}_n^{\dagger 2}(t)\hat{r}_n^2(t), 
\end{align}
where~$n$ runs over all bus states, and~$\hat{r}_n(t) = \hat{r}-\alpha_n(t)e^{-i\omega_dt}$. This interaction corresponds to a displaced self-Kerr nonlinearity of the NLR and will be shown to constrain the system dynamics to a low-energy manifold.
Moreover, we show in~\cref{sec:DD_scheme} how to trade this nonlinear interaction for two additional drives on the NLR.

Momentarily ignoring the effect of~$K_r$, the drive on the NLR grows a coherent state of amplitude~$\alpha_n$ satisfying 
\begin{align}
    i \dot\alpha_n(t) = (\delta+n\chi-i\kappa/2) \alpha_n(t) - \varepsilon(t), \label{eq:dalphandt}
\end{align}
which, because of the interaction~$\hat H_{br}$, is conditional on the bus state~$|n\rangle$. Here,~$\delta = \omega_r-\omega_d$ is the frequency detuning between the NLR and the drive, and~$\kappa$ is the single-photon loss rate of the NLR. Omitting the qubits, the Hamiltonian of~\cref{eq:Hlab} together with single-photon loss stabilizes join bus-NLR states of the form~$\ket{\psi_{n,k}(t)} = \ket{n}_b\ket{\alpha_n(t)e^{-i\omega_d t};k}_r$, where~$\ket{\alpha;k}_r = e^{\alpha \hat{r}^{\dagger}-\alpha^*\hat{r}} \ket{k}_r$ is the~$k$th Fock state displaced by an amplitude~$\alpha$. This can be more clearly seen by plotting the metapotential associated to the Hamiltonian~$\hat{H}$ with the qubit modes traced out and for~$K_r=0$, obtained by replacing the operator~$\hat r$ ($\hat r^\dag$) with the complex variable~$I+iQ$ ($I-iQ$). As illustrated in~\cref{fig:decoupling_mechanism}c), this metapotential has a single well corresponding to a stable point of the system and whose position in the~$I$-$Q$ plane is distinct for each bus state~$\ket{n}_b$. Moreover, because the latter states are associated with coherent states~$\ket{\alpha_n}_r$ that are disjoint in phase space, bus transitions are effectively suppressed.
If the system is energetically constrained to first state~$\ket{\psi_{n,0}(t)}$ of the metapotential wells, we recover~\cref{eq:nm_transition} where the matrix elements of the bus mode are exponentially suppressed with respect to the drive amplitude. The nonlinear interaction~$\hat H_{nlbr}$ of amplitude~$K_{r}$ plays the role of a self-Kerr nonlinearity within each well of the NLR metapotential. As a result, similarly as in the Kerr-cat qubit~\cite{Shruti2017}, this Kerr nonlinearity helps constrain the system's dynamics to the low-energy states of each of the metapotential wells. 

Rapid switching between the `on' and `off' states of the coupler is realized by taking advantage of the transitionless-quantum-driving (TQD) method to rapidly displace the NLR coherent state starting from vacuum~\cite{berry2009transitionless}. In the numerical simulations that are discussed below, we use the pulse shape
\begin{align}
    \varepsilon(t) = \left(\varepsilon_0(t) -  \frac{i\dot{\varepsilon}_0(t)}{\delta - i \kappa/2}\right)\Theta(\tau-t) + \varepsilon_0(\tau) \Theta(t-\tau),
    \label{eq:enveloppe}
\end{align}
where~$\varepsilon_0(t)$ is a smooth drive amplitude,~$\varepsilon_0(0)=0$, and~$\Theta(x)$ is the Heaviside step function. With this choice of drive envelope, the steady-state reached at time~$\tau$ takes the form
\begin{align}
    \bar{\alpha}_n \simeq \frac{\varepsilon_0(\tau)}{\delta+n\chi-i\kappa/2}
    \label{eq:alpha_n}
\end{align}
for each bus state~$\ket{n}_b$.

To avoid overlapping metapotential wells and strongly suppress the bus transitions $0\leftrightarrow n$, the system parameters are chosen such that~$\bar{\alpha}_0$ is large with respect to any other~$\bar{\alpha}_n$. This last requirement ensures that the bus ground state is well separated in energy from the NLR excitations, maximizing the exponential suppression of the two-qubit interaction. More precisely, this is achieved for~$\abs{\delta/\chi} \ll 1$ and~$\abs{\kappa/\chi} \ll 1$. We note that choosing the drive such as to make~$\bar{\alpha}_{n\neq0}$ large is also a valid strategy. However, we numerically find that increasing~$\bar{\alpha}_0$ performs better for small to moderate values of~$K_r$. Importantly, the TQD protocol is reversible and can be used to bring the coupler back to the `on' state by emptying the NLR in a time much faster than~$1/\kappa$. The details of this analysis are provided in~\cref{app:polaron_transformation}. 

\section{Numerical experiments \label{sec:numerics}}

\subsection{Suppression of bus transitions}\label{sec:Suppressing_boupler_Rabi} 

\begin{figure}[t!]
    \centering
    \includegraphics[width=0.9\textwidth]{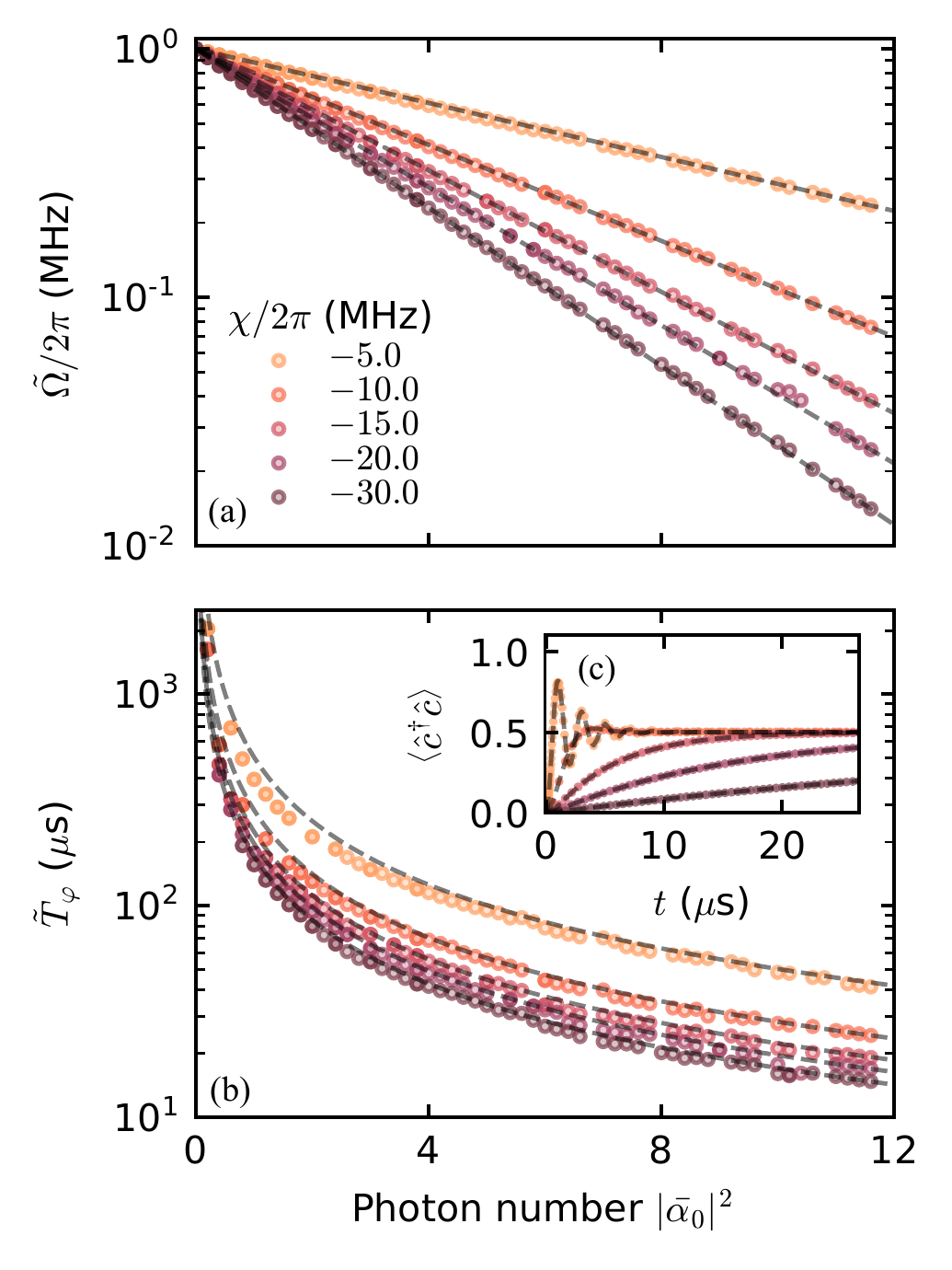}
    \caption{Renormalized bus a) Rabi frequency~$\tilde{\Omega}/2\pi$ and b) dephasing time~$\tilde{T}_{\varphi}$ under a resonant~$\Omega/2\pi = 1$ MHz microwave bus drive as a function of the photon number in the NLR,~$\abs{\bar{\alpha}_0}^2$. The results are obtained using a 5-ns-long TQD scheme with~$\kappa/2\pi = 100$ KHz,~$\delta/2\pi = -5$ MHz,~$K_b/2\pi = -300$ MHz. The cross-Kerr~$\chi/2\pi$ is varied in the range -5.0 to -30.0 MHz. Fits are done by comparing the time evolution to that of an effective two-level system, including both~$T_1$ and~$T_2$.  
    Dashed lines correspond to~\cref{eq:Rabi_renorm,eq:T2_renorm}. c) Example of the time-evolution traces that are fitted for~$\abs{\bar{\alpha}_0}^2 = 11$.}
    \label{fig:Rabi}
\end{figure}

We now turn to numerical simulations of the concepts presented in the previous section. To illustrate the working principle---the suppression of bus-state transitions in the presence of a drive on the NLR---we first simplify the setup by omitting the qubits. In lieu of the qubits, we add a drive term of the form 
\begin{equation}
    \hat{H}_{\mathrm{Rabi}}/\hbar = \Omega 
    \left(
        e^{-i \tilde{\omega}_b t}\hat{b}^{\dagger}+e^{i \tilde{\omega}_b t}\hat{b}
    \right),
\end{equation}
 where~$\Omega$ is the drive amplitude and the drive frequency~$\tilde{\omega}_b$ is set to ac-Stark shifted~$0-1$ transition frequency of the bus
\begin{align}
    \tilde{\omega}_b =  \omega_b + \frac{\delta(\delta+\chi) - (\kappa/2)^2}{\chi} \abs{\bar{\alpha}}^2,
\end{align}
with 
\begin{equation}
    \bar{\alpha} = \bar{\alpha}_1 - \bar{\alpha}_0 = - \frac{\chi}{\delta + \chi - i \kappa/2}\bar{\alpha}_0 \label{eq:alpha_bar}
\end{equation}
being the distance between the metapotential wells associated with the ground and first excited states of the bus. In the `on' state of the coupler, the resonant drive on the bus will result in Rabi oscillations between~$\ket{0}_b$ and~$\ket{1}_b$. In the `off' state, bus transitions, and therefore Rabi oscillations, are exponentially suppressed with the coherent state amplitude~$\bar{\alpha}_0$. 

Indeed, according to~\cref{eq:nm_transition}, we expect the Rabi frequency in the presence of a drive on the NLR to take the form 
\begin{align}
    \tilde{\Omega} \approx 
    \Omega \exp\left(-\abs{\bar{\alpha}}^2/2\right).
    \label{eq:Rabi_renorm}
\end{align}
Following~\cref{eq:alpha_bar},~$\abs{\bar{\alpha}}$ is bounded by~$\abs{\bar{\alpha}_0}$. Indeed,~$0 \leq \abs{\bar{\alpha}}<\abs{\bar{\alpha}_0}$ where the lower bound corresponds to~$\chi =0$ or~$\bar{\alpha}_0 = 0$, and the upper bound is reached for~$\abs{\chi}\rightarrow \infty$. As a result, increasing~$\abs{\chi}$ results in a stronger suppression of~$\tilde{\Omega}$.

\Cref{fig:Rabi} (a) shows the Rabi frequency~$\tilde{\Omega}$ obtained from numerical integration of the coupler master equation based on~\cref{eq:Hlab}. The result includes damping in the NLR but excludes decoherence in the bus, and it is computed for different equilibrium values of~$\abs{\bar{\alpha}_0}^2$ and cross-Kerr interactions~$\chi$. The data points are extracted from fits to the bus population~$\langle \hat b^\dag \hat b\rangle (t)$ with the bus and NLR initialized to the vacuum state, see panel (c). The numerical result (symbols) is in excellent agreement with~\cref{eq:Rabi_renorm} (dashed lines) and display the expected exponential suppression of the bus Rabi oscillations. This suppression becomes more significant for increasing cross-Kerr coupling~$\abs{\chi}/2\pi$ which is shown here ranging from 5 to 20 MHz. We note that these results are obtained for~$K_r = 0$. In the absence of the qubits ($g_j = 0$), choosing small~$\abs{\Omega/\delta}$ guarantees that the dynamics is mainly generated by states~$\ket{\psi_{n,0}}$ for which the exponential suppression of the Rabi frequency is maximized. Indeed, the states~$\ket{\psi_{0,k}}$ are separated in energy by~$\delta$ and thus, to prevent transitions to~$k\neq 0$ states during a~$1\rightarrow 0$ bus transition, we ideally require the matrix elements of Rabi drive Hamiltonian in the state basis~$\ket{\psi_{n,k}}$ to be small relative to $\delta$, i.e.~$\abs{\Omega\matrixel{\psi_{0,k}}{\hat b}{\psi_{1,0}}/k\delta} = \abs{\tilde{\Omega}\bar{\alpha}^k/k\delta\sqrt{k!}}\ll 1$ for~$k\neq 0$.

In the presence of single-photon loss in the NLR, the distinct coherent states associated to the different bus states lead to bus dephasing. This originates from the `which-bus-state' information that is carried by the lost photons, something that is akin to measurement induced-dephasing in the dispersive readout of circuit QED~\cite{Gambetta2006}. With~$T_\varphi$ denoting the bare bus dephasing time, the coherence time in the presence of the NLR drive takes the form
\begin{align}
    \tilde{T}_\varphi \approx \left(\frac{1}{T_\varphi} + \frac{\kappa}{2}\abs{\bar{\alpha}}^2\right)^{-1}.
    \label{eq:T2_renorm}
\end{align}
\cref{fig:Rabi} (b) shows this dephasing time extracted from the numerical simulations including~$\kappa \neq 0$ [symbols]. Similarly to the previous case, we find excellent agreement with the analytical expression (dashed lines). To isolate the effects of NLR dissipation on the system, we have omitted intrinsic relaxation and dephasing of the bus mode. A key observation is that, while transitions between the bus states are suppressed exponentially with~$\bar{\alpha}$, dephasing only increases polynomially with this quantity. Moreover, we demonstrate below that the dephasing induced on the bus mode does not percolate to the qubits.

\subsection{Suppression of two-qubit interactions} 

Having numerically confirmed that suppressing the bus transitions by driving the NLR is possible, we now reincorporate the qubits to the model and explore the two-qubit decoupling. In particular, we characterize the hybridization between the qubit and bus modes as a function of the NLR drive parameters and demonstrate how spurious two-qubit couplings, such as the~$ZZ$ interaction, are exponentially suppressed. 

\subsubsection{Polaron transformation}

Analyzing the underlying physics of the model Hamiltonian is made easier by applying a rotating frame transformation and polaron-like unitaries that displace the NLR mode conditionally on the state of the bus. Acting with~\cref{eq:polaron_transformation} on~\cref{eq:Hlab}, the transformed Hamiltonian takes the form (see~\cref{app:polaron_transformation})
\begin{equation}
    \hat{H}^P = \sum_{j=1}^2\hat{H}_{j}^P + \hat{H}_{br}^P + \hat{H}_{\kappa}^P + \hat{H}_g^P, \label{eq:Hp}
\end{equation}
with
\begin{align}
    & \hat{H}_{j}^P/\hbar = \tilde{\Delta}_j  \hat{q}_j^{\dagger}\hat{q}_j + \frac{K_j}{2}\hat{q}_j^{\dagger 2}\hat{q}_j^2, \\
    & \hat{H}_{br}^P/\hbar = \left(\delta + \chi\hat{b}^{\dagger}\hat{b}\right)\hat{r}^{\dagger}\hat{r} + \frac{\tilde{K}_{b}}{2}\hat{b}^{\dagger 2}\hat{b}^2 + \frac{K_{r}}{2}\hat{r}^{\dagger 2}\hat{r}^2, \\
    & \hat{H}_{\kappa}^P/\hbar = \frac{i\kappa}{2}\sum_{n}\left(\alpha_n\hat{r}^{\dagger}-\alpha_n^*\hat{r}\right)\ketbra{n}{n}_b, \\
    & \hat{H}_g^P/\hbar = \sum_{j,n}g_j \hat{q}_j^{\dagger} e^{i\phi_n}\hat{D}_{n,r} \sqrt{n+1}\ketbra{n}{n+1}_b + \mathrm{h.c.}, \label{eq:HgP}
\end{align}
where the NLR decay rate~$\kappa$ appears in the displacement transformation according to~\cref{eq:dalphandt}. In~$\hat{H}_g^P$ we have defined the bus-state-dependent phases
\begin{align}
   \phi_n = \frac{\alpha_{n+1}^*\alpha_{n}-\alpha_{n}^*\alpha_{n+1}}{2i}, \label{eq:phin}
\end{align}
the NLR displacement operators 
\begin{align}
    \hat{D}_{n,r} = e^{\left(\alpha_{n+1}-\alpha_{n}\right)\hat{r}^{\dagger}-\left(\alpha_{n+1}^*-\alpha_{n}^*\right)\hat{r}}.
\end{align}
and the ac-Stark shifted qubit-bus detunings and bus anharmonicity
\begin{align}
    & \tilde{\Delta}_j = \omega_{j} -  \omega_b - \delta \abs{\alpha_0}^2 +  (\delta+\chi) \abs{\alpha_1}^2, \label{eq:DeltaTilde}\\
    & \tilde{K}_b = K_b - \delta \abs{\alpha_0}^2 + 2(\delta +\chi)\abs{\alpha_1}^2 - (\delta+2\chi) \abs{\alpha_2}^2,
\end{align}
respectively. In what follows, we neglect the transients to focus on times where the polaronic states are fully grown with~$\alpha_n=\bar{\alpha}_n$ as defined in~\cref{eq:alpha_n}.

In the polaron frame Hamiltonian~$\hat H^P$, all modes are described by a Kerr nonlinear oscillator Hamiltonian. Moreover, the interaction between the qubits and the bus,~$\hat{H}_g^P$, reflects the fact that transitions in the bus are accompanied by displacements of the NLR field. Damping is described by the usual Lindblad master equation in the polaron frame, derived in~\cref{sec:PframeME}. In particular, the NLR photon-loss Lindblad operator~$\hat{L}_r=\sqrt{\kappa}\hat{r}$ in the laboratory frame transforms to~$\hat{L}_r^P = \sqrt{\kappa}\left(\hat{r} + \sum_n  \bar{\alpha}_n\ketbra{n}{n}_b\right)$ in the polaron frame. The action of both~$\hat{H}_{\kappa}^P$ and $\hat{L}_r^P$ ensures the stabilization of the vacuum state in the NLR for all~$\ket{n}_b$. Under the rotating-wave approximation, the Lindblad dynamics of the system can be further reduced to the effective Hamiltonian~$\hat{H}^P-\hat{H}_{\kappa}^P$ together with the two Lindblad operators~$\sqrt{\kappa}\hat{r}$ and~$\sqrt{\kappa} \sum_n \bar{\alpha}_n\ketbra{n}{n}_b$. By assuming the dynamics to be restrained to the ground and first-excited states of the bus, the latter operator takes the simpler form (c.f.~\cref{app:polaron_transformation}) 
\begin{align}
    \hat{L}_b^P = \sqrt{\frac{\kappa \abs{\bar{\alpha}}^2}{4}} \left(\ketbra{1}{1}_b - \ketbra{0}{0}_b\right). \label{eq:LbP}
\end{align}
This result is in agreement with the expression for the bus dephasing time derived in~\cref{eq:T2_renorm}.

\begin{figure}[t!]
    \centering
    \includegraphics[width=\textwidth]{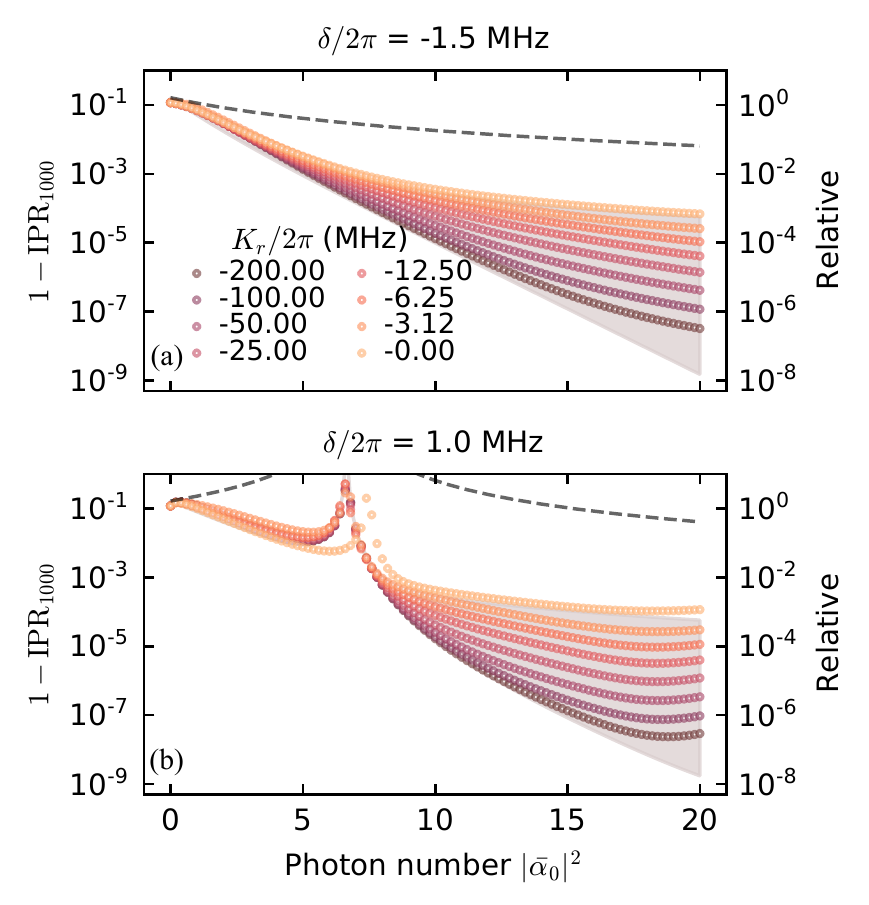}
    \caption{Suppression of~$1-\mathrm{IPR}_{1000}$ with respect to the NLR photon number~$\abs{\bar{\alpha}_0}^2$. The data points are obtained from numerical diagonalization with a fixed drive amplitude in the NLR. Analogous plots for~$1-IPR_{0100}$ can be found in~\cref{app:IPR}. Dashed lines correspond to the same system but with the NLR undriven and the bus frequency tuned to the ac-Stark shifted frequency found in the driven system. The gray regions are bounded by the analytical estimates in~\cref{eq:ipr1000_analytical} for~$K_r\rightarrow \infty$ and~$K_r = 0$. Here~$(\omega_{1}-\omega_b)/2\pi = 7.0$ MHz,~$(\omega_{2}-\omega_b)/2\pi = 14.0$ MHz,~$K_{1}/2\pi = K_{2}/2\pi = -300.0$ MHz,~$\chi/2\pi = -20.0$ MHz, and~$g/2\pi = 2.0$ MHz. 
    }
    \label{fig:IPR}
\end{figure}

\subsubsection{Inverse participation ratio}

A useful quantity to further characterize the exponential suppression of the two-qubit interactions is the mode hybridization between the qubits and the bus in the `off' state. To quantify this effect we make use of the inverse participation ratio (IPR)~\cite{DiVincenzo2020,Evers2008}, which here takes the form 
\begin{align}
    \mathrm{IPR}_{\mu} :=  \frac{\sum_{\nu}\abs{\braket{\psi_{h,\mu}}{\psi_{b,\nu}}}^4}{ \abs{\braket{\psi_{h,\mu}}{\psi_{h,\mu}}}^2},
    \label{eq:IPR}
\end{align}
where~$\ket{\psi_{b,\nu}}$ and~$\ket{\psi_{h,\nu}}$ the bare ($g_j=0$) and hybridized ($g_j\neq0$) eigenstates of the full system. The IPR is a measure of how localized the wavefunctions of the circuit modes are with respect to the bare modes, and it ranges from~$1/4$ (maximally delocalized states) to~$1$ (maximally localized states) for our system with four modes. 

In the dispersive regime where the qubits are largely detuned from the bus mode, analytical expressions for the IPR can be obtained by eliminating~$\hat{H}_{g}^P$ in~$\hat{H}^P$ using a Schrieffer-Wolff transformation. For the bare states~$\ket{\psi_{b,\nu}}$ with~$\nu \in \{1000, 0100, 1100\}$ and where the state indexing corresponds to~$(Q_1, Q_2, B, R)$, we obtain in~\cref{app:IPR} 
\begin{align}
    & \mathrm{IPR}_{1000} \approx 1 - 2 e^{-\abs{\bar{\alpha}}^2} \left(\frac{g_1}{\tilde{\Delta}_1}\right)^2 {}_4 F_4 \left(\boldsymbol{p}_1; 1 + \boldsymbol{p}_1; \abs{\bar{\alpha}}^2\right),\label{eq:ipr1000_analytical}\\
    & \mathrm{IPR}_{0100} \approx 1 - 2 e^{-\abs{\bar{\alpha}}^2} \left(\frac{g_2}{\tilde{\Delta}_2}\right)^2 {}_4 F_4 \left(\boldsymbol{p}_2; 1 + \boldsymbol{p}_2; \abs{\bar{\alpha}}^2\right)  ,\label{eq:ipr0100_analytical}
\end{align}
and~$\mathrm{IPR}_{1100} \approx \mathrm{IPR}_{1000}+\mathrm{IPR}_{0100}-1$, where~$\bar{\alpha} = \bar{\alpha}_1-\bar{\alpha}_0$,~${}_pF_q$ is the generalized hypergeometric function,~$\boldsymbol{p}_j = \begin{pmatrix} p_{j-} & p_{j-} & p_{j+} & p_{j+} \end{pmatrix}$ with~$p_{j\pm} = \beta [1 \pm   \sqrt{ 1  + \smash[b]{2\tilde{\Delta}_j \beta^{-2}/K_r}}]$, and~$\beta = (\delta+\chi)/K_r - 1/2$.
The expressions for the IPR show that the degree of mode hybridization decreases with increasing detuning. More importantly, we also see that hybridization is exponentially suppressed with increasing photon population of the NLR mode. At large photon number, virtual transitions to higher-energy states within the NLR's metapotential wells can impact the level of exponential suppression, something that is represented by the contribution from the hypergeometric function. These higher energy transitions can, however, be prevented by increasing~$\abs{\delta+\chi}$ and~$\abs{K_r}$.

To verify these observations, we investigate the quantity~$1-\mathrm{IPR}_{1000}$ as a function of the photon number~$|\bar{\alpha}_0|^2$ in the NLR by exact diagonalization of the Hamiltonian of~\cref{eq:Hp} for~$\kappa = 0$ (see~\cref{fig:IPR}). Different colors correspond to different values of the nonlinearity~$K_r$. Focusing first on panel (a), obtained for~$\delta/2\pi = -1.5$~MHz, the anticipated suppression of the hybridization with increasing~$|\bar{\alpha}_0|^2$ is clearly observed, together with the slowdown of that trend for larger~$\abs{\bar{\alpha}_0} \propto \abs{\bar{\alpha}}$. 

The shaded region is plotted using the analytical expression in~\cref{eq:ipr1000_analytical} for~$K_r$ in the range~$\abs{K_r} \rightarrow \infty$ to~$K_r = 0$. In the former limit, the dynamics is constrained to the low-lying polaronic states, i.e.~the NLR is constrained to the displaced Fock states~$\ket{0}$ and~$\ket{1}$ in the laboratory frame, and the exponential suppression persists for large~$\abs{\bar{\alpha}_0}$. As a comparison, the dashed line is obtained from the usual dispersive factor~$(g_i/\tilde\Delta_i)^2$ taking into account the change in qubits-bus detuning due to the ac-Stark shift and which corresponds to usual qubit-bus-qubit couplers without the driven NLR.
The very strong suppression of~$1-\mathrm{IPR}$ observed in~\cref{fig:IPR} (a) for our coupler design has an important consequence: because of the very small hybridization of the qubit eigenstates, all real and virtual qubit-qubit interactions mediated by the coupler are exponentially suppressed in amplitude; see~\cref{app:qubit_qubit_interactions} for details. Analogous plots for~$1-\mathrm{IPR}_{0100}$ can be found in~\cref{app:IPR}. 

\Cref{fig:IPR}(b) also shows~$1-\mathrm{IPR}_{1000}$ as a function of the number of photons~$|\bar{\alpha}_0|^2$ in the NLR but now for a positive detuning of~$\delta/2\pi = 1$~MHz. In this situation, we observe a divergence in the IPR associated with a resonance in the ac-Stark shifted detunings~$\tilde{\Delta}_1$. As discussed in~\cref{app:parameter_regimes}, this resonance can be understood from the poles of the generalized hypergeometric function appearing in~\cref{eq:ipr1000_analytical,eq:ipr0100_analytical}, which correspond to frequency collisions with higher energy levels of the NLR. For negative detunings~$\delta$, these collisions are avoided and the suppression is monotonic with photon number. On the other hand, choosing~$\delta>0$ results in a nonmonotonic IPR but this can lead to a stronger suppression of unwanted interactions. See~\cref{app:parameter_regimes} for a detailed discussion of these frequency collisions and  how to take advantage of them. 

We conclude this section with a discussion about the optimal sweep rate across resonances to minimize their impact on the qubits' dynamics. Frequency collisions in~\cref{fig:IPR,fig:IPR_2} correspond to pairs of states in the uncoupled system that are brought into resonance during a sweep of~$\bar{\alpha}_0$. To understand the impact of these collisions, let us consider the toy-model Hamiltonian $\hat{H}_\lambda(t) = \lambda e^{i \int_0^t dt' \Delta_\lambda(t') }\hat{q}_1^\dagger\hat{b}+\mathrm{h.c.}$ between~$Q_1$ and the bus in the interaction picture. Here,~$\Delta_\lambda(t)$ corresponds to the instantaneous detuning between these systems including ac-Stark shifts, and $\lambda$ is an effective coupling strength. In particular, we focus on the simplest case where~$\Delta_\lambda = \Delta_1 - \delta \nu t$, with $\nu = \abs{\bar{\alpha}_0(t)}^2/t$ constant. The approximate Bogoliubov angle~$\theta$ between $Q_1$ and the bus at the end of the sweep can be estimated using a first-order Magnus expansion of the time-evolution operator, according to~$\theta =  \abs{-i\int_0^{t_{\rm crit} + \tau} dt' H_\lambda(t')}$, where $t_{\rm crit} = \Delta_1/\delta \nu$ is the time at which the resonance occurs for $\delta/\Delta_1 > 0$, and $t_{\rm crit}+\tau$ is the total time of the sweep. More precisely, we arrive at
\begin{equation}
    \theta = \abs{\frac{\lambda}{\Delta_1}\frac{\Delta_1}{\sqrt{\delta \nu/2}} \frac{\sqrt{\pi}}{2}\mathrm{Erf}\left(\frac{e^{i\pi/4}\Delta_\lambda(t')}{\sqrt{\delta\nu/2}}\right)\Big|_{t'=0}^{t_{\rm crit}+\tau}}, \label{eq:Lambda}
\end{equation}
where $\lambda/\Delta_1$ is a measure of the dispersive coupling between the qubit and the bus away from the resonance. Two limiting cases can be studied for $\Delta_\lambda(t_{\rm crit}+\tau)/\Delta_\lambda(0)<0$ and therefore $\delta/\Delta_1>0$. For a slow sweep ($\nu \ll \abs{\Delta_1^2/\delta}$) and~$\abs{\Delta_\lambda(t)}\gg \abs{\lambda}$ at $t=0$ and $t=t_ {\rm crit}+\tau$, we derive the asymptotic expression
\begin{equation}
    \theta \approx \abs{\frac{\sqrt{\pi}\lambda}{\sqrt{\delta\nu/2}}},
\end{equation}
which is minimized for~$\nu \gg \abs{2\pi \lambda^2/\delta}$. If $\delta \sim \lambda$, we recover the intuitive result $\nu \gg \abs{\lambda}$, where the sweep rate must exceed the gap set by the coupling strength $\lambda$. For faster sweeps ($\nu \gg \abs{\Delta_1^2/\delta}$), which are the ideal case, $\theta \approx 0$. We stress the the sweeping rate $\nu$ is limited by  the cross-Kerr interaction strength~$\chi$ [c.f. \cref{app:polaron_transformation}] that impacts the TQD protocol. In the toy model $\hat H_\lambda$, the envelope is $\varepsilon_0(t) = \delta \sqrt{\nu t}$. It follows that $d^k \varepsilon_0(t)/dt^k = (1/2)^{(k)}\nu^k \varepsilon_0(t)/\bar{\alpha}_0^{2k}(t)$ must be much smaller in magnitude than $|\chi^{k+1}|$. In other words, $\abs{2\chi^2 \bar{\alpha}_0(t_{\rm crit}+\tau)/\delta} \gg \nu \gg \abs{\lambda}$. This implies that sweeping rate across the resonances should ideally exceed the coupling strengths $g_j$ and be smaller than $2\chi^2/\delta$, where the upper limit avoids non-adiabatic errors from an imperfect TDQ protocol.

\begin{figure}[t!]
    \centering
    \includegraphics[width=\textwidth]{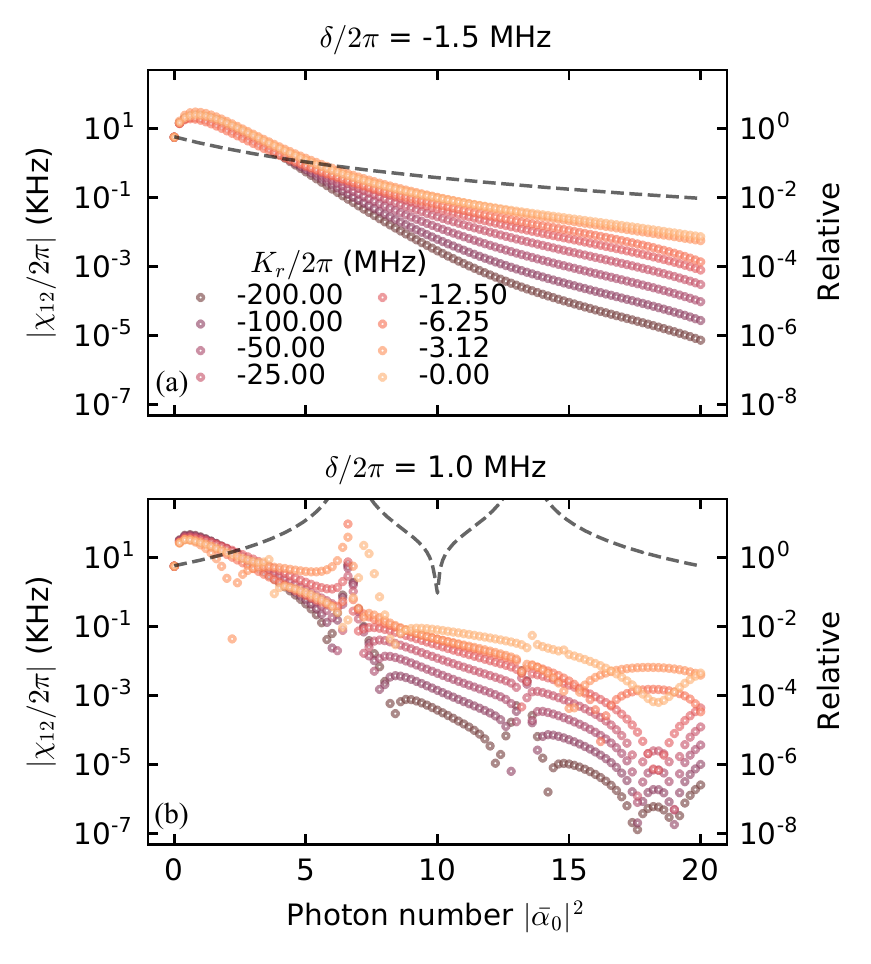}
    \caption{Suppression of the~$ZZ$ interaction~$\chi_{12}$ between the qubits as a function of the NLR photon number~$\abs{\bar{\alpha}_0}^2$. Data points correspond to numerical diagonalization of the system Hamiltonian with a fixed drive amplitude in the NLR. Dashed lines correspond to~\cref{eq:chi_AC}, i.e.~the same system but with the NLR undriven and the bus frequency tuned to the ac-Stark shifted frequency found in the driven system. Here~$(\omega_{1}-\omega_b)/2\pi = 7.0$ MHz,~$(\omega_{2}-\omega_b)/2\pi = 14.0$ MHz,~$K_{1}/2\pi = K_{2}/2\pi = -300.0$ MHz,~$\chi/2\pi = -20.0$ MHz, and~$g/2\pi = 2.0$ MHz. 
    }\label{fig:ZZ}
\end{figure}

\subsubsection{Suppression of spurious interactions}

We now analyze how the proposed coupler help to suppress the spurious cross-Kerr coupling between the qubits, which is given by
\begin{align}
    & \chi_{12} = \omega_{1100} - \omega_{1000} - \omega_{0100} + \omega_{0000}, 
\end{align}
where $\omega_{\mu} = \langle\psi_{h,\mu_{\star}}|\hat{H}^P-\hat{H}_{\kappa}^P|\psi_{h,\mu_{\star}}\rangle$ is the energy associated with the two-qubit eigenstate~$\ket{\psi_{h,\mu_{\star}}}$ with maximal overlap with the bare state~$\ket{\psi_{h,\mu}}$, i.e.~$\mu_{\star} = \mathrm{arg max}_{\nu}\abs{\braket{\psi_{h,\nu}}{\psi_{b,\mu}}}^2$. 

\cref{fig:ZZ} shows~$|\chi_{12}|$ obtained from numerical diagonalization of~$\hat{H}^P-\hat{H}_{\kappa}^P$ as a function of~$\abs{\bar{\alpha}_0}^2$, for different values of~$K_r$ (symbols). As a comparison, the dashed line shows~$|\chi_{12}|$ resulting only from the change in detuning between the qubits and the bus due to the ac-Stark shift, and computed using the perturbative expression
\begin{align}
    \chi_{12}^{\rm ac} =
    \frac{1}{6}\frac{g_1^2}{\tilde{\Delta}_1}\frac{g_2^2}{\tilde{\Delta}_2}\left(\frac{1}{\tilde\Delta_1}+\frac{1}{\tilde\Delta_2}\right), \label{eq:chi_AC}
\end{align}
valid for~$|\tilde{\Delta}_j/K_j| \ll 1$ and~$|\tilde{\Delta}_j/\tilde{K}_{b}| \ll 1$. The latter is obtained from a Magnus expansion to fourth order in the coupling strengths [c.f. \cref{app:qubit_qubit_interactions}]. The two resonances observed in the dashed line and the numerical data in panel (b) correspond to~$\tilde{\Delta}_j=0$. 

As first noticed for~$1-\mathrm{IPR}$ in~\cref{fig:IPR}, the suppression of~$\chi_{12}$ is monotonic with photon number for negative detunings~$\delta$ [panel (a)], while some nonmonotonic features appear at positive detuning where the suppression is also stronger [panel (b)]. See~\cref{app:parameter_regimes} for a discussion of the origin of these features.

We also note the presence of more features in~\cref{fig:ZZ}b) for~$\chi_{12}$ than in~\cref{fig:IPR}b) for the IPR. The first two dominant peaks in~\cref{fig:ZZ}b) result from accidental resonances between each qubit and the bus, i.e.~$\tilde{\Delta}_j = 0$, in agreement with the peaks observed for the IPR in~\cref{fig:IPR} and~\cref{fig:IPR_2}. Additional features in~\cref{fig:ZZ} not present in \cref{fig:IPR,fig:IPR_2} result from frequency collisions with higher energy levels in the system, activated by the ac-Stark shifts in the bus. 

Importantly, the suppression of~$1-$IPR and the resulting reduction of the spurious cross-Kerr coupling does not require fine-tuning of the circuit or drive parameters. Indeed, as illustrated in~\cref{fig:IPR,fig:ZZ}, strong suppression is observed for different choices of circuit parameters including~$\delta$ and~$K_r$. It is also worth emphasizing that all real and virtual interactions are suppressed by this scheme. This fact is in stark contrast to other approaches where cancellation of two-qubit interactions is realized only for a precise value of a control parameter and where residual virtual interactions such as~$\chi_{12}$ remain present~\cite{Plourde2020,Yu2020,Poletto2021}. 

Finally, we note that it is possible to combine our coupler with other approaches for suppressing spurious interactions, for instance by using qubits with opposite sign anharmonicities. 

\subsection{Bus-induced qubit dephasing}

At the origin of the suppression of unwanted interaction are the disjoint bus-state dependent coherent states of the driven NLR. A photon lost from the NLR carries the `which-bus-state' information and leads to dephasing of the bus state. Because there exists hybridization between the bus and qubit modes, this mechanism can introduce additional qubit dephasing. However, as shown in more details in~\cref{app:polaron_transformation}, we find that this is not an important contribution to qubit dephasing. Indeed, by expressing~\cref{eq:LbP} in the hybridized eigenbasis, the dephasing rate of the first qubit is given by
\begin{equation}\label{eq:dephasing_rates_1}
    \begin{split}
        \gamma_{\varphi,1} 
        &= \frac{\kappa \abs{\bar{\alpha}}^2}{2} \left(\sum_{k=0}^{\infty}\abs{\braket{\psi_{h,1000}}{\psi_{b,001k}}}^2\right)^2\\
        &\approx \frac{\kappa \abs{\bar{\alpha}}^2}{2}  \frac{1-\mathrm{IPR}_{1000}}{2},
    \end{split}
\end{equation}
where the second line follows from a Schrieffer-Wolff transformation [c.f.~\cref{app:measurement_dephasing}]. The expression above was obtained with a rotating-wave approximation, which is valid for~$|\gamma_{\varphi,1}/\tilde{\Delta}_1|\ll 1$. An expression for the second qubit is obtained by simply replacing the subscript~$1000$ by~$0100$.

Similarly to measurement-induced dephasing~\cite{Gambetta2006}, the prefactor of~\cref{eq:dephasing_rates_1} scales with the photon number~$|\alpha|^2$ in the NLR. However, because~$1-\mathrm{IPR}$ is exponentially suppressed with increasing~$|\bar{\alpha}|^2$, the qubit dephasing rate can be made negligible in the `off' state of the coupler. \Cref{app:measurement_dephasing} also compares~\cref{eq:dephasing_rates_1} versus photon number against the result obtained from numerical diagonalization of~$\hat{H}^P-\hat{H}_{\kappa}^P$. As with the suppression of unwanted~$ZZ$ interactions, the reason for this negligibly small dephasing rate is the very low hybridization of the qubits' eigenstates with the bus and NLR modes.

\section{Effective parametric modulation}\label{sec:DD_scheme}

In the previous sections, we have seen that large nonlinear interaction amplitudes~$K_r$ help in the suppression of the unwanted interactions in the `off' state of the coupler. Here, we explore an alternative strategy that relies on a two-tone drive on the NLR. Moreover, because the nonlinearity is not needed in this case, the NLR can be taken to be a linear resonator (LR). This might also simplify the experimental realization of these ideas.

Our starting point is again the Hamiltonian of~\cref{eq:Hlab} where we now take~$K_r = 0$ and introduce the following additional drive on the LR
\begin{align}\label{eq:Hdd}
    \hat{H}_{\rm DD} = -\frac{i\lambda \omega_m}{2\bar{\alpha}^*} \left[e^{-i\left(\omega_r -\omega_m\right)t}-e^{-i\left(\omega_r +\omega_m\right)t}\right]\hat r^\dagger + \mathrm{h.c.},
\end{align}
where~$\lambda$ is a real-valued amplitude and the frequency~$\omega_m$ is assumed here to be much larger in magnitude than the cross-Kerr interaction~$\chi$. With this additional two-tone drive on the LR, the steady-state bus-dependent coherent state~\cref{eq:alpha_n} becomes
\begin{align}
    \bar{\alpha}_n \rightarrow \bar{\alpha}_n - i\lambda \cos(\omega_m t)/\bar{\alpha}^*,\label{eq:alphanmod}
\end{align}
where~$\lambda$ plays the role of the amplitude of a modulation around the steady-state value~$\bar{\alpha}_n$. Crucially, this modulation changes the phase~$\phi_n$ that specifies the bus-state-dependent displacements Hamiltonian~$\hat H_g^P$ in~\cref{eq:HgP}, which can now be written as
\begin{align}
    & \phi_n(t) =  \bar{\phi}_n - \lambda \mathrm{Re}\left[\frac{\bar{\alpha}_{n+1}-\bar{\alpha}_n}{\bar{\alpha}}\right] \cos(\omega_m t), \\
    & \bar{\phi}_n = \frac{\bar{\alpha}_{n+1}^*\bar{\alpha}_{n}-\bar{\alpha}_{n}^*\bar{\alpha}_{n+1}}{2i}.
 \end{align}
 Moreover, the qubit-bus detunings transform to~$\tilde{\Delta}_j = \tilde{\bar{\Delta}}_{j} + \tilde{\Delta}_j^t$ where 
 \begin{align}
     & \tilde{\bar{\Delta}}_{j} = \omega_{j} -  \omega_b - \delta \abs{\bar{\alpha}_0}^2  +  (\delta+\chi) \abs{\bar{\alpha}_1}^2 + \frac{\chi \lambda^2}{2\abs{\bar{\alpha}}^2}, \\
     & \tilde{\Delta}_j^t = -2\chi\lambda \mathrm{Im}\left[\frac{\bar{\alpha}_1}{\bar{\alpha}}\right]  \cos(\omega_mt) + \frac{\chi \lambda^2}{2\abs{\bar{\alpha}}^2}\cos(2\omega_mt).
 \end{align}
An additional rotating frame transformation such as to remove the time-dependence of the qubit-bus detunings, leads to the following approximation for~$\hat H_g^P$
\begin{align}\label{eq:HgPp}
    & \hat{H}_g^{P\prime} = \sum_{j,n}g_j \hat{q}_j^{\dagger} e^{i\phi_{n}^\prime(t)}\hat{D}_{n,r} \sqrt{n+1}\ketbra{n}{n+1}_b + \mathrm{h.c.},
\end{align}
where we have introduced~$\phi_{n}^\prime(t) = \phi_n(t) + \int_0^t dt \tilde{\Delta}_j^t$. Except for the now time-dependent phase~$\phi_n^\prime(t)$,~\cref{eq:HgPp} as the same form as~\cref{eq:HgP}. 

The role of the time-dependent phase~$\phi_n^\prime(t)$ can be understood by using the Jacobi-Anger expansion
\begin{align}\label{eq:Jacobi-Anger}
    e^{i\phi_n^\prime} = e^{i\bar{\phi}_n} \sum_{s=-\infty}^{+\infty} (-i)^s J_s\left(\lambda\,\mathrm{Re}\left[\frac{\bar{\alpha}_{n+1}-\bar{\alpha}_n}{\bar{\alpha}}\right] \right)e^{is\omega_m t},
\end{align}
where~$J_s(z)$ is the~$s$th Bessel function of the first kind, and where we considered negligible $\chi/\omega_m$ for simplicity. Because the bus mode is ideally only virtually excited at all times, our goal is to dominantly suppress the~$0\leftrightarrow 1$ transition in the bus. To achieve this, we adjust the amplitude~$\lambda$ to reach a zero of~$J_0$ in~\cref{eq:Jacobi-Anger}, noticing that~$\bar{\alpha}=\bar{\alpha}_1-\bar{\alpha}_0$. Higher harmonics of~\cref{eq:Jacobi-Anger} oscillate rapidly for~$\omega_m \gg \abs{\chi}, |\tilde{\bar{\Delta}}_{j}|$, and result in a lower bound on the suppression of the two-qubit interactions (c.f.~\cref{app:DD}). 

\begin{figure}[t!]
    \centering
    \includegraphics[width=0.9\textwidth]{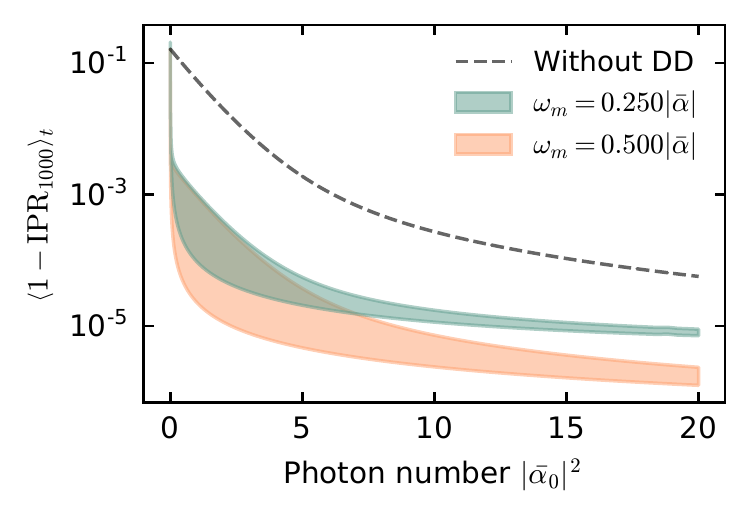}
    \caption{Time-averaged inverse participation ratio against photon number using a two-tone drive on the LR ($K_r = 0$) with frequencies~$\omega_r\pm \omega_m$ and amplitudes~$\omega_m \lambda /\bar{\alpha}^*$. The shaded regions are bounded by~$\lambda = \lambda_0$ where~$J_0(\lambda_0) = 0$ and a 10\% error on~$\lambda_0$. 
    The other parameters are~$\delta /2\pi=-1.0$ MHz,~$\chi/2\pi = -20.0$ MHz,~$(\omega_{1}-\omega_b)/2\pi = 7.0$ MHz,~$(\omega_{2}-\omega_b)/2\pi = 14.0$ MHz,~$K_{1}/2\pi = K_{2}/2\pi = -300.0$ MHz, and~$g/2\pi = 2.0$ MHz.~$\left<1-\mathrm{IPR}_{0100}\right>_t$ can be found in~\cref{app:DD}.
    }
    \label{fig:IPR_DD}
\end{figure}

To understand how the two proposed implementation mechanisms compare to each other, we compute in~\cref{app:IPR} the IPR using a time-dependent Schrieffer-Wolff transformation to find 
\begin{align}
    \nonumber 1 - \mathrm{IPR}_{1000} \approx  2 g_1^2 e^{-\abs{\bar{\alpha}}^2} \sum_{s_1,s_2=-\infty}^{+\infty} i ^{s_2-s_1}e^{i(s_2-s_1)\omega_m t} & \\
    \times J_{s_1}(\lambda)J_{s_2}(\lambda)\frac{{}_4 F_4 \left(\boldsymbol{p}_{1s_1s_2}; 1 + \boldsymbol{p}_{1s_1s_2}; \abs{\bar{\alpha}}^2\right)}{\left(\tilde{\Delta}_1+s_1\omega_m\right)\left(\tilde{\Delta}_1+s_2\omega_m\right)},& \label{eq:IPRmod}
\end{align}
where~$\boldsymbol{p}_{js_1s_2} = \begin{pmatrix} p_{js_1-} & p_{js_2-} & p_{js_1+} & p_{js_2+} \end{pmatrix}$ with~$p_{js\pm} = \beta [1 \pm   \sqrt{ 1  + \smash[b]{2(\tilde{\Delta}_j+s\omega_m) \beta^{-2}/K_r}}]$ and~$\beta = (\delta+\chi)/K_r - 1/2$.We observe that~\cref{eq:IPRmod} is reminiscent of~\cref{eq:ipr1000_analytical}, and a similar expression for~$\mathrm{IPR}_{0100}$ can be derived. In the large~$\omega_m \gg |\tilde{\Delta}_1|$ limit, the dominant contribution to~\cref{eq:IPRmod} arises from the term with~$s_1=s_2=0$, which is canceled by adjusting~$\lambda$ to reach a zero of~$J_0$. Importantly, because of the already suppressed interactions, there is no need for a very fine adjustment of~$\lambda$. The time-averaged IPR according to~\cref{eq:IPRmod} is illustrated in~\cref{fig:IPR_DD}, where we take~$\omega_m = \omega_0 \abs{\bar{\alpha}}$ such that the  drive amplitude in~\cref{eq:Hdd} is independent of~$\abs{\bar{\alpha}}$. The dashed line corresponds to the absence of dynamical decoupling. The shaded regions correspond to~$\pm$ 10\% error bounds on the drive amplitude. We observe a strong suppression of 1-IPR$_{1000}$ and of the drive amplitude sensitivity. As discussed further in~\cref{app:DD}, we note that the asymptotic behavior of the suppression is polynomial in~$\bar{\alpha}$. With the very large suppression of the 1-IPR that is observed in~\cref{fig:IPR_DD}, this is a small price to pay when trading the nonlinearity~$K_r$ for an additional drive. We finally note that the suppression can be further enhanced with the help of a longitudinal drive in the LR (c.f.~\cref{app:DD}).

\section{Superconducting circuit implementation \label{sec:implementation}}

\begin{figure*}[t!]
    \centering
    \includegraphics[width=\textwidth]{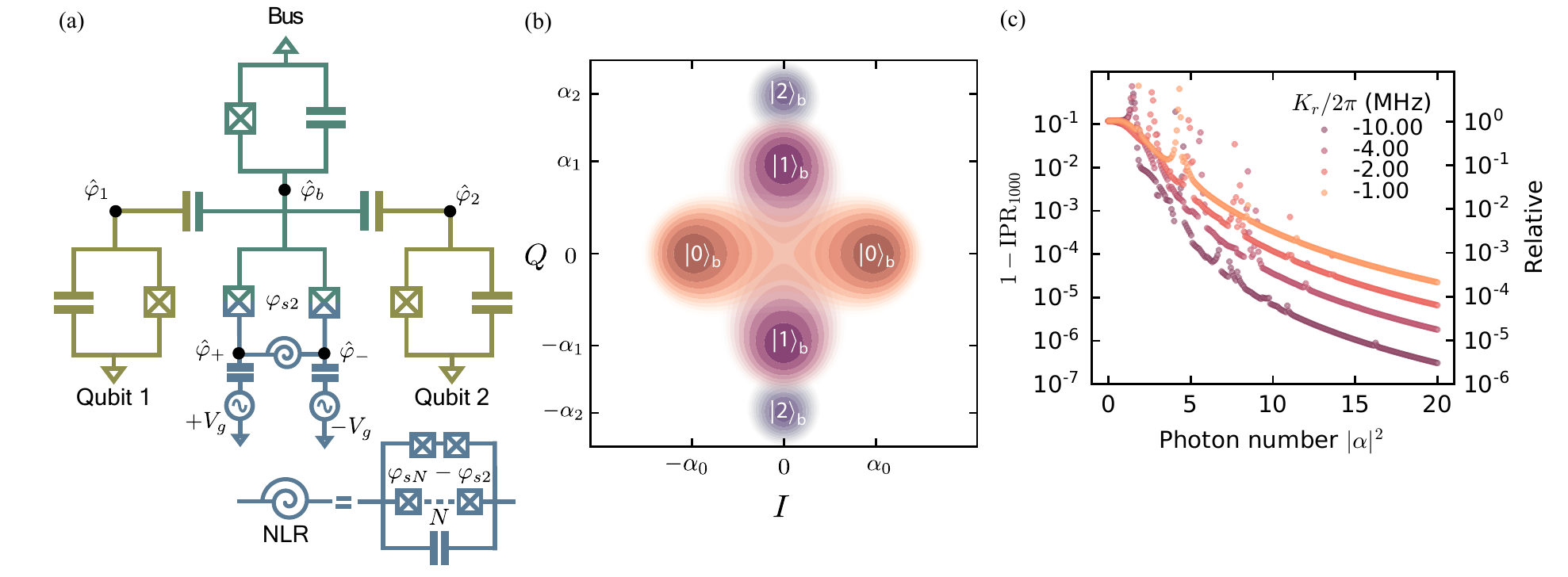}
    \caption{Superconducting circuit implementation. a) Circuit design. The qubits and bus modes are implemented using transmon qubits;~$\hat\varphi_1$,~$\hat\varphi_2$ and~$\hat\varphi_b$ are the phase operators of $Q_1$, $Q_2$ and the bus modes respectively. Here~$2\hat\varphi_\pm = -\hat\varphi_b \pm \left(\hat\varphi_r-\varphi_\ell\right)$ where~$\varphi_\ell$ is a real-valued scalar to be defined. A SNAIL-like element, representing the NLR mode with phase operator~$\hat\varphi_r$, is linearly driven by a voltage source~$V_g$.~$\Phi_0\varphi_{s2}/2\pi$ and~$\Phi_0\left(\varphi_{sN}-\varphi_{s2}\right)/2\pi$ are two external fluxes that control the bus-NLR interaction. b) Metapotential of the NLR for each of the bus states~$\ket{0}_b$ (orange),~$\ket{1}_b$ (purple) and~$\ket{2}_b$ (blue). Here~$\delta/2\pi = -5.0$ MHz,~$\chi/2\pi = -5.0$ MHz and~$K_r/2\pi = -10.0$ MHz. c)~$1-\mathrm{IPR}_{1000}$ (and~$1-\mathrm{IPR}_{0100}$ can be found in~\cref{app:circuit}) estimated by numerical diagonalization of the full system using the effective Hamiltonian in~\cref{eq:HKerrCat} with~$\delta /2\pi=1$ KHz,~$\chi/2\pi = -5.0$ MHz,~$(\omega_{1}-\omega_b)/2\pi = 7.0$ MHz,~$(\omega_{2}-\omega_b)/2\pi = 14.0$ MHz,~$K_{1}/2\pi = K_{2}/2\pi = -300.0$ MHz, and~$g/2\pi = 2.0$ MHz. 
    }
    \label{fig:circuit_implementation}
\end{figure*}

In this section, we introduce a superconducting quantum circuit realizing our coupler. To approach the model Hamiltonian of~\cref{eq:Hlab}, we draw inspiration from the Kerr-cat qubit, which exploits the bifurcation physics of driven Josephson-based devices~\cite{Blais2017,Grimm2020}. A simplification based on the idea of dynamical decoupling presented in~\cref{sec:DD_scheme} is also discussed.

\subsection{Kerr-cat-based circuit model}

\Cref{fig:circuit_implementation}(a) shows a possible circuit realization of our coupler with two transmon qubits interacting through a transmon-like device playing the role of bus mode. The latter mode is connected to a driven nonlinear circuit representing the NLR and consisting of a loop formed by two symmetrical Josephson junction and a SNAIL-like element which incorporates an array of $N\sim 3$ junctions~\cite{Devoret2017}. 
Omitting the qubits, the Hamiltonian of the circuit reads
\begin{align}
    & \hat{H} = \hat{H}_b + \hat{H}_r + \hat{H}_{br}, \label{eq:Hcircuit}
\end{align}
where 
\begin{equation}\label{eq:Hb}    
    \begin{split}
        \hat{H}_b &= 4 E_{C_b}\hat{n}_b^2 -E_{J_b}\cos\left(\hat{\varphi}_b\right) \approx \omega_b \hat b^{\dagger}\hat b + \frac{K_b}{2}\hat b^{\dagger 2}\hat b^2, 
    \end{split}
\end{equation}
is the bus Hamiltonian, which we treat as a weakly nonlinear oscillator of frequency~$\omega_b = \sqrt{8 E_{C_b}E_{J_b}} - E_{C_b}$ and anharmonicity~$K_b = - E_{C_b}$. We define the the phase operators of the two modes across the SNAIL-like element as~$2\hat\varphi_\pm =- \hat\varphi_b \pm \left(\hat\varphi_r-\varphi_\ell\right)$ where~$\hat \varphi_b$ ($\hat \varphi_r$) is the phase operator of the bus (NLR) and~$\varphi_\ell$ is a real-valued scalar determined from the minimization of the potential energy of the circuit. We consider two external flux biases:~$\varphi_{s2}$ in the three-node loop and~$\varphi_{sN}-\varphi_{s2}$ in the SNAIL-like circuits. Here,~$\varphi_{s2}$ ($\varphi_{sN}$) shifts the cosine potential of the two junctions (N junctions) in the SNAIL-like circuit. Moreover,~$\hat\varphi_1$ ($\hat \varphi_2$) is the phase operator of $Q_1$ ($Q_2$). The NLR and bus-NLR Hamiltonians take the form
\begin{align}
    \nonumber & \hat{H}_r  =  4 E_{C_r}\hat{n}_r^2 -  NE_{J_N}\cos\left(\frac{\hat{\varphi}_r-\varphi_{sN}}{N}\right) + 2\epsilon(t)\hat{n}_r \\
    & \quad - 2  E_{J_\ell}   \cos\left(\frac{\hat{\varphi}_r-\varphi_\ell}{2}\right) -  2E_{J_2}\cos\left( \frac{\hat{\varphi}_r-\varphi_{s2}}{2}\right),\\
    & \hat{H}_{br} =  - 2  E_{J_\ell}   \left[\cos\left(\frac{3\hat{\varphi}_b}{2}\right)-1\right] \cos\left(\frac{\hat{\varphi}_r-\varphi_\ell}{2}\right).
\end{align}
In these expressions,~$E_{C_r}$ is the NLR charging energy,~$E_{J_\ell}$ the Josephson energy of the NLR's symmetrical junctions,~$E_{J_2}$ the Josephson energy of each of the two small NLR's junctions, and~$E_{J_N}$ the Josephson energy of each of the~$N$ large junctions in the array. Moreover,~$\epsilon(t)$ is the amplitude of the drive of frequency~$2(\omega_r-\delta)$ on the NLR, where~$\omega_r$ is the frequency of the undriven NLR. 

The idea is to stabilize cat states in the NLR with amplitudes that depend on the bus photon number. Just as in the simplified model discussed in the previous section, transitions between bus states are associated to displacements in the NLR. An advantage of this propoposed realization is that the large anharmonicity in the NLR is now determined by the size of the cat state.
To this end, we follow \textcite{Devoret2017} by choosing the external fluxes and Josehpson energy such as to obtain a cubic nonlinearity of the form~$\hat b^{\dagger}\hat b \left(\hat r^{\dagger} + \hat r\right)^3$ in~$\hat{H}_{br}$. In the presence of a linear drive on the NLR, the cubic nonlinearity leads to a nearly resonant, bus-photon-number-dependent two-photon drive in the NLR. The Kerr nonlinearity in the NLR can then stabilize bus-photon-number-dependent cat-states.

More precisely, we take~$E_{J_2}=E_{J_\ell}$,~$\varphi_\ell = \pi-2\zeta$ and~$\varphi_{s2} = -\pi - 2\zeta$ where~$\zeta$ is a parameter to be defined. With these choices, we have
\begin{align}
    \hat H_r\approx 4 E_{C_r}\hat{n}_r^2 -  NE_{J_N}\cos\left(\frac{\hat{\varphi}_r-\varphi_{sN}}{N}\right) + 2\epsilon(t) \hat{n}_r,
\end{align}
and
\begin{align}
    \nonumber \hat{H}_{br} \approx & \frac{9\pi z_b  E_{J_\ell} }{4} \left(2\hat{b}^{\dagger}\hat{b}+1\right)\\
    & \times \left[\cos\zeta\sin\left(\frac{\hat{\varphi}_r}{2}\right)+\sin\zeta \cos\left(\frac{\hat{\varphi}_r}{2}\right)\right], 
\end{align}
where~$z_{b(r)}=Z_{b(r)}/R_Q$ is the reduced impedance of the bus (NLR) mode with~$R_Q\simeq 6.5\,\mathrm{k}\Omega$ the resistance quantum. In~$\hat{H}_{br}$, the sine and cosine terms that depend on~$\hat\varphi_r$ are key for implementing the bus-photon-number-dependent cubic nonlinearity~$\hat b^{\dagger}\hat b \left(\hat r^{\dagger} + \hat r\right)^3$ and the cross-Kerr interaction~$\hat b^{\dagger}\hat b \hat{r}^\dagger\hat{r}$. In presence of the drive~$\epsilon(t) = \epsilon_0 \sin[2(\omega_r-\delta)t]$ where, for simplicity, we take a constant drive amplitude~$\epsilon_0$, our next step is to apply a displacement transformation~$\hat D[\xi(t)]$ on the NLR mode to eliminate the drive term. To achieve this, we take~$- i\dot{\xi} + \omega_r \xi + i \epsilon(t)/\sqrt{\pi z_r} = 0$ or, equivalently,~$\sqrt{\pi z_r}\xi(t) \approx -\Omega \cos[2(\omega_r-\delta)t]+ (i \Omega/2) \sin[2(\omega_r-\delta)t]$ with the displacement amplitude~$\Omega = 2\epsilon_0/3\omega_r$. By doing so, we obtain the displaced Hamiltonian~$\hat H^D = \hat H_b + \hat{H}_r^D + \hat{H}_{br}^D$ with  
\begin{align}
    & \hat{H}_r^D \approx \omega_r \hat{r}^{\dagger}\hat{r} + \frac{K_r}{2}\hat{r}^{\dagger 2}\hat{r}^2 +  \frac{\lambda_r}{2} \hat{r}^{\dagger 2} e^{-i2(\omega_r-\delta)t} + \text{h.c.}\\
    & \hat{H}_{br}^D \approx \chi\left(\hat{b}^{\dagger}\hat{b}+\frac{1}{2}\right) \hat{r}^{\dagger}\hat{r} \nonumber\\
    & \phantom{\hat{H}_{br}^D =} + \frac{\lambda_\ell }{2}\left(2\hat{b}^{\dagger}\hat{b}+1\right)\hat{r}^{\dagger 2}e^{-i2(\omega_r-\delta)t} + \text{h.c.}
\end{align}
for small~$\varphi_{sN}/N$. The above expressions are valid for small reduced mode impedance~$\pi z_r \approx \sqrt{2NE_{C_r}/E_{J_N}}$ assume the rotating-wave approximation. We have introduced the NLR frequency~$\omega_r = \sqrt{8 E_{C_r} E_{J_N}/N} - E_{C_r}$, the self-Kerr anharmonicity~$K_r  = - E_{C_r}/N^2$, the two-photon drive amplitude
\begin{align}
    \lambda_r = -\frac{(\pi z_r)^{3/2} \Omega  E_{J_N}  \varphi_{sN}}{2 N^3},
\end{align}
the cross-Kerr interaction amplitude 
\begin{align}
    \chi = - \frac{9\pi z_b \pi z_r  E_{J_\ell}\sin(\zeta)}{8},
\end{align}
and the bus-number dependent two-photon drive amplitude 
\begin{align}
    \lambda_\ell = \frac{9\pi z_b (\pi z_r)^{3/2} \Omega E_{J_\ell}\cos(\zeta)}{64}.
\end{align}
We moreover set~$\lambda_r = - 2 \lambda_\ell$ by a proper choice of the flux biases.

Finally, in a doubly rotating frame at~$\omega_r-\delta$ for~$\hat{r}$ and at~$\omega_b$ for~$\hat{b}$, the displaced Hamiltonian reads 
\begin{equation}\label{eq:HKerrCat}
    \begin{split}
    & \hat{H}'^D  \approx  \frac{K_b}{2}\hat{b}^{\dagger 2}\hat{b}^2 - \frac{K_r \alpha^4}{2}\left(2\hat{b}^{\dagger}\hat{b}-1\right)^2 \\
    & \phantom{\hat{H}^D} + \left(\delta + \chi/2 + \chi \hat{b}^{\dagger}\hat{b}\right)\hat{r}^{\dagger}\hat{r} \\
    & \phantom{\hat{H}^D} + \frac{K_r}{2}\left[\hat{r}^{\dagger 2} + \alpha^2\left(2\hat{b}^{\dagger}\hat{b}-1\right)\right]\left[\hat{r}^{2} + \alpha^2\left(2\hat{b}^{\dagger}\hat{b}-1\right)\right], 
    \end{split}
\end{equation}
where~$\alpha^2 = \lambda_\ell/K_r$. 

The metapotential associated to this Hamiltonian is illustrated in~\cref{fig:circuit_implementation}(b). As in~\cref{fig:decoupling_mechanism} for the simplified system, the different Fock states of the bus mode lead to displaced wells in the~$I$-$Q$ plane. However, because of the combination of the Kerr nonlinearity and the engineered two-photon drive, each Fock state is associated two metapotential wells~\cite{Blais2017}. The central idea of blocking bus-state transition by entangling those state to coherent states in the NLR is, however, unchanged. This is confirmed in~\cref{fig:circuit_implementation}(c) which shows 1-IPR as a function of photon number. Apart from additional resonance which can easily be avoided, the overall behavior is the one expected: we see a an exponential reduction of the bus-state hybridization with NLR photon number, as originally predicted by the model of~\cref{eq:Hlab}.

Furthermore, it is useful to note that the ac-Stark shift on the bus frequency vanishes in this model, i.e. the ground- and first-excited states of the bus are both shifted in energy by~$K_r\alpha^4/2$. The bus-NRL entanglement is therefore entirely responsible for the exponential suppression observed in~\cref{fig:circuit_implementation}c).

\subsection{Harmonic model with parametric modulation}

In~\cref{sec:DD_scheme} we have seen how it is possible to trade the large nonlinear interaction between the bus and the NLR by additional drives. Here, we show how this idea can be realized without modifications to the circuit of~\cref{fig:circuit_implementation}. For this second approach, we take~$\varphi_\ell = 0$,~$\varphi_{s2} = 2 \pi$,~$E_{J_2}=\lambda E_{J_\ell}$ with~$\lambda = 1 - (3/2)^2 \pi z_b/2$, and~$\varphi_{sN} \ \mathrm{mod} \  2\pi N = 0$. With these parameter choices, the circuit Hamiltonian can now be written as
\begin{align}
    & \hat{H}_b = 4 E_{C_b}\hat{n}_b^2 -E_{J_b}\cos\left(\hat{\varphi}_b\right) - 2  E_{J_\ell}   \cos\left(\frac{3\hat{\varphi}_b}{2}\right), \\
    & \hat{H}_r  =  4 E_{C_r}\hat{n}_r^2 -  NE_{J_N}\cos\left(\frac{\hat{\varphi}_r}{N}\right) + 2\epsilon(t)\hat{n}_r. \\ 
    & \hat{H}_{br} =  - 2  E_{J_\ell}   \left[\cos\left(\frac{3\hat{\varphi}_b}{2}\right)-\lambda\right] \left[\cos\left(\frac{\hat{\varphi}_r}{2}\right)-1\right],
\end{align}
We note that the reduced mode impedance of the NLR,~$\pi z_r \approx \sqrt{2 N E_{C_r}/E_{J_N}}$, needs to be made small to prevent the drive on the NLR from resulting in appreciable nonlinear terms due to the cosine potentials and the nearly resonant two-photon and cubic terms. More precisely, we take~$1/\sqrt{\pi z_r}$ to be much larger than any displacement in the NLR field associated with the bus Fock states~$n\neq 0$, and small compared to~$N/\sqrt{\pi z_r}$ for~$n=0$. 

As above, we treat the bus and NLR as weakly nonlinear oscillators. Importantly, for~$n=0$, we find that~$\hat{H}_{br}\approx 0$ and~$\hat{H}_r$ is approximately harmonic despite a large displacement in the NLR field. Moreover, for~$n\neq 0$, ~$\hat{H}_{br}$ effectively implements a large cross-Kerr interaction that strongly reduces displacements of the NLR field by rendering the linear drive of the NLR off-resonant. Particularly, we find that 
\begin{equation}
    \begin{split}
        &\hat{H} \approx \sum_{\nu=b,r} \left( \omega_\nu \hat{\nu}^{\dagger}\hat{\nu} + \frac{K_\nu}{2}\hat{\nu}^{\dagger 2} \hat{\nu}^2 \right) + \chi \hat{b}^{\dagger}\hat{b}\hat{r}^{\dagger}\hat{r} \\
        & \phantom{\hat{H} \approx} + i\Omega(t) \left(\hat{r}^{\dagger}-\hat{r}\right).
    \end{split}
\end{equation}
In this expression, we have defined the frequencies~$\omega_\nu  \approx \sqrt{8 E_{C_\nu} E_{L_\nu}} - E_{C_\nu}$, the reduced mode impedances~$\pi z_\nu \approx \sqrt{2 E_{C_\nu}/E_{L_\nu}}$, the inductive energies~$E_{L_b} = E_{J_b} + 2(3/2)^2 E_{J_\ell}$ and~$E_{L_r} = E_{J_N}/N$, the anharmonicities~$K_b = -[E_{J_b}+2 (3/2)^4 E_{J_\ell}] E_{C_b}/E_{L_b}$ and~$K_r = E_{C_r}/N^2$, and the linear drive amplitude~$\Omega(t) =  \epsilon(t)/2\sqrt{\pi z_r}$. 

We also largely reduce the NLR's anharmonicity~$K_r$ by choosing a small~$E_{C_r}$ and large~$N$. It is possible with this model to observe the exponential suppression of two-qubit coupling by choosing~$\epsilon(t)$ to be nearly resonant with the NLR. 

As discussed in~\cref{sec:DD_scheme}, with a reduced anharmonicity in the NLR an additional drive, which we choose to be of the form
\begin{align}
    \Omega(t) = 2\delta\bar{\alpha}_0 \cos\left[(\omega_r-\delta) t\right] + \frac{\omega_m\lambda}{\bar{\alpha}}\sum_{\nu = \pm }\nu\sin\left[(\omega_m+\nu \omega_r)t\right]
\end{align}
can serve as a complementary mechanism to suppress interactions. In the limit~$\abs{\delta/\chi}\ll 1$,~$\abs{\chi/\omega_m}\ll 1$ and~$\abs{\omega_m/\omega_r}\ll 1$, we find that the bus-state dependent displacements take the form 
\begin{align}
    \alpha_n(t) & \approx \frac{\delta\bar{\alpha}_0}{\delta + n\chi} -\frac{i\lambda}{\bar{\alpha}}\cos(\omega_m t),
\end{align}
in agreement with~\cref{eq:alphanmod}. From this point on, the result of~\cref{sec:DD_scheme} follows. 

Finally, we emphasize that even though a cross-Kerr type interaction between the bus and the NLR could, in principle, be implemented using a dispersive coupling~\cite{Blais2020}, the dispersive Hamiltonian is invalid at large photon numbers and yields virtual qubit-qubit interactions through the driven NLR. We also note that a discussion of the leading effects of stray couplings can be found in~\cref{app:circuit}.

\section{Conclusion} 

We introduced a two-qubit coupler with an exponential on-off ratio, realized by connecting a pair of qubits to a bus mode complemented by a driven ancillary resonator. The cross-Kerr interaction between the bus and the driven resonator results in a displacement of the resonator's field that is conditional on the bus state. Because the displaced resonator states have negligible overlap, bus-state transitions are suppressed exponentially in the amplitude of the drive. In turn, because two-qubit interactions are mediated by bus transitions, the two-qubit coupling also results strongly suppressed, leading to a high on-off ratio. As a clear demonstration of this mechanism, we have shown how the inverse participation ratio, which is a measure of qubit-bus hybridization, and the spurious cross-Kerr between the qubits are exponentially reduced with the number of photons in the resonator mode. We also proposed two complementary superconducting quantum circuit implementations of our coupler. 

The strong reduction in two-qubit couplings demonstrated here can be advantageous in multiqubit processors, where spectator qubits and long-range qubit-qubit interactions can have detrimental effects~\cite{DiVincenzo2020}. For the same reason, the proposed approach can be particularly useful in all-microwave frequency-fixed qubits architectures with interactions mediated by frequency-fixed buses. We hope that the mechanisms explored here will pave the way to a new generation of coupling schemes for diverse platforms. 

Finally, we note that possible improvements to the coupler include squeezing the resonator mode to further reduce the overlaps between the resonator states associated with distinct bus states, and extending the ancillary system to multiple modes such that the exponential suppression is now with respect to multiple modes.

\section*{Acknowledgments}

We thank Ross Shillito and Jens Koch for useful discussions. This work was undertaken in part thanks to funding from NSERC, the Canada First Research Excellence Fundm the Minist\`ere de l'\'economie et de l’innovation du Qu\'ebec and the U.S. Army Research Office Grant No. W911NF-18-1-0411

\bibliography{refs.bib}

\newpage

\appendix

\onecolumngrid

\section{Polaron transformation\label{app:polaron_transformation}}

To gain intuition about the underlying physics of the model, it is useful to move to a frame defined by the time-dependent polaron transformation
\begin{align}
    \hat{P}(t) = e^{-i\int_0^t d\tau [\omega_b+\Delta_{\rm ac}(\tau)] \left(\hat{b}^{\dagger}\hat{b}+\sum_j\hat{q}_j^{\dagger}\hat{q}_j\right)}e^{-i\omega_dt \hat{r}^{\dagger}\hat{r}} \sum_{n=0}^{\infty}\hat{D}_r[\alpha_n(t)]\otimes |n\rangle\langle n|_b, & \label{eq:polaron_transformation},
\end{align}
where~$|n\rangle\langle n|_b$ is the projection operator associated with the eigenstate~$|n\rangle$ of the bus mode, and~$\Delta_{\rm ac}$ is an ac-Stark shift that will be defined below. The displacements~$\{\alpha_n(t)\}$ are determined from the damped classical equation of the NLR [c.f.~\cref{eq:dalphandt}] that result from the bilinear Hamiltonian terms only.

\subsection{Transformed Hamiltonian}

Under~\cref{eq:polaron_transformation} the Hamiltonian~\cref{eq:Hlab} transforms to~$\hat{H}^{\rm P} = \hat P^{\dagger} \hat{H} \hat P - i \hat P^{\dagger} \dot {\hat P}$, where
\begin{align}
    & \hat{H}^P = \sum_{j=1}^2\hat{H}_j^P + \hat{H}_{br}^P + \hat{H}_{\kappa}^P + \hat{H}_g^P,\label{eq:Hp_app}\\
    & \hat{H}_j^P =(\omega_{j}-\omega_b-\Delta_{\rm ac}) \hat{q}_j^{\dagger}\hat{q}_j + \frac{K_j}{2}\hat{q}_j^{\dagger 2}\hat{q}_j^2, \\
    & \hat{H}_{br}^P = \left(\delta + \chi\hat{b}^{\dagger}\hat{b}\right)\hat{r}^{\dagger}\hat{r} + \frac{K_b}{2}\hat{b}^{\dagger 2}\hat{b}^2 + \frac{K_{r}}{2}\hat{r}^{\dagger 2}\hat{r}^2  + \sum_{n}\Delta_{\rm ac,n} \ketbra{n}{n}_b, \\
    & \hat{H}_{\kappa}^P = \frac{i\kappa}{2}\sum_{n}\left(\alpha_n\hat{r}^{\dagger}-\alpha_n^*\hat{r}\right)\ketbra{n}{n}_b, \\
    & \hat{H}_g^P = \sum_{j,n}g_j\left(\hat{q}_j^{\dagger} e^{i\phi_n}\hat{D}_{n,r} \sqrt{n+1}\ketbra{n}{n+1}_b + \mathrm{h.c.}\right),
\end{align}
with~$i2\phi_n = \alpha_{n+1}^*\alpha_{n}-\alpha_{n}^*\alpha_{n+1}$,~$\hat{D}_{n,r} = \hat{D}_r\left(\alpha_{n+1}-\alpha_{n}\right)$,~$\hat{D}_r(\alpha) = e^{\alpha \hat{r}^{\dagger}-\alpha^*\hat{r}}$ is the displacement operator in the NLR, and 
\begin{align}
    & \Delta_{\rm ac,n} = \delta \abs{\alpha_0}^2 - (\delta+n\chi)\abs{\alpha_n}^2 - n \Delta_{\rm ac}, \\
    & \Delta_{\rm ac} = \delta \abs{\alpha_0}^2-(\delta+\chi)\abs{\alpha_1}^2.
\end{align}
As we only consider Jaynes-Cummings-type interactions, the transformed Hamiltonian can be reduced to the form in~\cref{eq:Hp}. In addition, driving the NLR mode introduces an ac-Stark shift~$\Delta_{\rm ac}^C(t) = \Delta_1^C(t) - \Delta_0^C(t)$ of the bus frequency given by
\begin{align}
    \Delta_{\rm ac}^C(t) = \delta \abs{\alpha_0(t)}^2-(\delta+\chi)\abs{\alpha_1(t)}^2.
    \label{eq: ac stark}
\end{align}

\subsection{Transformed Master equation}
\label{sec:PframeME}

The full system dynamics can be described by the Lindblad Master equation formalism,
\begin{align}
    \dot{\hat{\rho} }= - i \comm{\hat H}{ \hat \rho} + \sum_j \hat L_j \hat \rho \hat L_j^{\dagger} - \frac{1}{2}\acomm{\hat L_j^{\dagger}\hat L_j}{\hat \rho}
\end{align}
where~$\hat\rho$ is the density matrix of the system,~$\hat H$ is the Hamiltonian and~$\hat L_j$ are the collapse operators. Under the transformation~\cref{eq:polaron_transformation} the density matrix transforms as~$\hat \rho = \hat P \hat \rho^P \hat P^\dagger~$. It follows that 
\begin{align}
    \dot{\hat{\rho} }^P= - i \comm{\hat H^P}{ \hat \rho^P} + \sum_j \hat L_j^P \hat \rho^P \hat L_j^{P \dagger} - \frac{1}{2}\acomm{\hat L_j^{P \dagger}\hat L_j^P}{\hat \rho^P}, \label{eq:rhoP}
\end{align}
where we defined the transformed collapse operators~$\hat L_j^P = \hat P^\dagger \hat L_j \hat P$. As examples, the collapse operators can take the form~$\hat L_r = \sqrt{\kappa }\hat r$,~$\hat L_{\nu,\gamma} = \sqrt{\gamma_\nu} \ketbra{0}{1}_\nu$ and~$\hat L_{\nu,\varphi} = \sqrt{\gamma_{\varphi,\nu}/2} \left(\ketbra{1}{1}_\nu-\ketbra{0}{0}_\nu\right)$ for~$\nu = \{b, 1, 2\}$. We find that qubit collapse operators as well as~$\hat L_{b,\varphi}$ are unchanged under the polaron transformation. However, we have
\begin{align}
    & \hat L_{b,\gamma}^P = \sqrt{\gamma_b} e^{-i\phi_0}e^{-i\int_0^td\tau \left[\omega_b+\Delta_{\rm ac}(\tau)\right]} \hat{D}_{0,r}^\dagger \ketbra{0}{1}_b,\\
    & \hat L_r^P = \sqrt{\kappa} \left(\hat r + \sum_n \alpha_n \ketbra{n}{n}_b\right).
\end{align}
from where it follows that~$\hat{L}_{b,\gamma}^P$ is exponentially suppressed because of the displacement operator in the NLR. Irrespective of this observation, it is worth nothing that~$\hat{L}_{b,\gamma}^P$ does not prevent the formation of the polaronic states in the coupler and therefore does not hinder the proposed protocol. In addition,~$\hat{L}_r^P$ corresponds to measurement-induced dephasing in the coupler. It is possible to further simplify the master equation within the rotating-wave approximation to 
\begin{align}
    \dot{\hat{\rho} }^P =  - i \comm{\hat H^P-\hat H_\kappa^P}{ \hat \rho^P} + \sum_j \hat {\tilde{L}}_j^P \hat \rho^P \hat {\tilde{L}}_j^{P \dagger} - \frac{1}{2}\acomm{\hat {\tilde{L}}_j^{P \dagger}\hat {\tilde{L}}_j^P}{\hat \rho^P},  
\end{align}
where~$\hat{\tilde{L}}_j^P=\hat{L}_j^P$, except~$\hat{\tilde{L}}_r^P = \sqrt{\kappa}\hat{r}$ and we define a new collapse operator~$\tilde{\hat{L}}_{b,\gamma_{\alpha}} = \sqrt{\kappa} \sum_n \alpha_n \ketbra{n}{n}_b$ which captures measurement-induced dephasing in the bus.

\subsection{Transitionless driving\label{app:TQD}}

Large conditional displacements in the NLR (`off' state) can be prepared by controlling the phase of the envelope in~\cref{eq:enveloppe} in time. The same mechanism makes it possible to empty the NLR quickly (`on' state). Imperfections in the envelope lead to deviations in the intended displacements, which we characterize in this section. Using~\cref{eq:dalphandt} we find the displacements 
\begin{align}
    \alpha_n(t) = i\int_0^t dz \ \varepsilon(z) e^{i(\delta-i\kappa/2 +n\chi) (z- t)} + \alpha_n(0) e^{-i(\delta-i\kappa/2 +n\chi) t}.
\end{align}
Extending the envelope in~\cref{eq:enveloppe} to include both switching-off and switching-on events
\begin{align}
    \nonumber \varepsilon(t) =& \left(\varepsilon_0(t) -  \frac{i\dot{\varepsilon}_0(t)}{\delta - i \kappa/2}\right)\Theta(\tau-t) + \varepsilon_0(\tau) \Theta(t-\tau)\Theta(T+\tau-t) \\
    & + \left(\varepsilon_0(\tau)-\varepsilon_0(t-T-\tau) +  \frac{i\dot{\varepsilon}_0(t-T-\tau)}{\delta - i \kappa/2}\right)\Theta(t-T-\tau)\Theta(T+2\tau-t),
\end{align}
where~$\tau$ is the ramping time to switch off/on the device and~$T$ is the time during which the drive is on, we find that 
\begin{align}
    \begin{split}
    \alpha_n(t) = & \frac{1}{\delta - i \kappa/2 +n\chi}\begin{cases}0, & t = 0 \\ \varepsilon_0(t), & 0\leq t \leq \tau \\ \varepsilon_0(\tau), & \tau\leq t\leq \tau+T \\ \varepsilon_0(\tau)-\varepsilon_0 (t-\tau-T), &\tau + T \leq t \leq 2\tau + T \\ 0, & t\leq 2\tau + T \end{cases}  \\
    & -\frac{i n\chi}{\delta-i\kappa/2} \left(\int_0^tdz\Theta(\tau-t)-\int_{T+\tau}^tdz \Theta(T+2\tau-t)\right) \frac{d\varepsilon_0(z)}{dz}\frac{e^{i(\delta-i\kappa/2 +n\chi) (z - t)}}{(-i)(\delta - i \kappa/2 +n\chi)} \\
    = & \frac{1}{\delta - i \kappa/2 +n\chi}\begin{cases}0, & t=0 \\ \varepsilon_0(t), & 0\leq t \leq \tau \\ \varepsilon_0(\tau), & \tau\leq t\leq \tau+T \\ \varepsilon_0(\tau)-\varepsilon_0 (t-\tau-T), &\tau + T \leq t \leq 2\tau + T \\ 0, & t\leq 2\tau + T \end{cases}  \\  
    & -\frac{ n\chi}{\delta-i\kappa/2} \sum_{k=1}^{\infty}\frac{d^k\varepsilon_0(z)}{dz^k}  \left.\frac{e^{i(\delta-i\kappa/2 +n\chi) (z - x)}}{(-i)^k(\delta - i \kappa/2 +n\chi)^{k+1}}\right|_{z=0}^{z=x}\left( \delta(t-x)\Theta(\tau-t)- \delta(t-\tau-T-x) \Theta(2\tau+T-t)\right),
    \end{split} \label{eq:alphan_TQD}
\end{align}
where we assumed~$\alpha_n(0)=0$. It follows that, for 
\begin{align}
    \abs{\frac{d^k\varepsilon_0(z)}{dz^k} \Big |_{z = 0}}, \ \abs{\frac{d^k\varepsilon_0(z)}{dz^k} \Big |_{z = \tau}}  \ll \abs{\delta-i\kappa/2}\abs{\chi}^k,
\end{align}
\cref{eq:alphan_TQD} simplifies to~\cref{eq:alpha_n} for~$\tau \leq t\leq \tau +T$, and vanishes for~$t\geq 2\tau +T$. Ultimately, the derivatives of the pulse at the endpoints of the ramp would contribute the most to deviations in the conditional displacements~$\alpha_n$. If one has perfect control over the pulse envelope the ramping time can be made arbitrarily small but limitations could arise from pulse imperfections. To see this, we define the perturbed envelope 
\begin{align}
    \varepsilon(t) \rightarrow \varepsilon(t) + \varepsilon_{\rm err}(t),
\end{align}
where~$\varepsilon_{\rm err}(t)$ is a small time-dependent perturbation. Using integration by parts, we find that
\begin{align}
    \alpha_n(t) \rightarrow \alpha_n(t) + i \sum_{k=0}^\infty \left.\frac{d^k\varepsilon_{\rm eff}(z)}{dz^k}\frac{e^{i(\delta-i\kappa/2 +n\chi) (z - t)}}{(-i)^k \left(\delta-i\kappa/2+n\chi\right)^k}\right|_0^t.
\end{align}
The effects of non-adiabatic errors are quantified by the ratio between the time derivatives of the drive envelope at the endpoints and powers of~$\abs{\delta-i\kappa/2 +n\chi}$. These errors result in time-dependent fluctuations of~$\alpha_n(t)$, which can change the conditional displacements~$\alpha_n$ and the ac-Stark shift of the bus. When switching on, TQD errors could result in residual photons in the NLR and ac-Stark shifts in the bus that can affect two-qubit interactions in the `on' state. It is therefore desirable to have a reset scheme for the NLR. 

\section{Numerical experiments and analytical estimates\label{app:numerics}}

In this section, we provide details regarding the numerical simulations, the derivations for the analytical estimates associated with the inverse participation ratio and the spurious two-qubit interactions. We also report additional numerical results for the inverse participation ratio and measurement-induced dephasing.

\subsection{Rabi drive experiment\label{app:rabi}}

\begin{figure}[h!]
    \centering
    \includegraphics[width=0.5\textwidth]{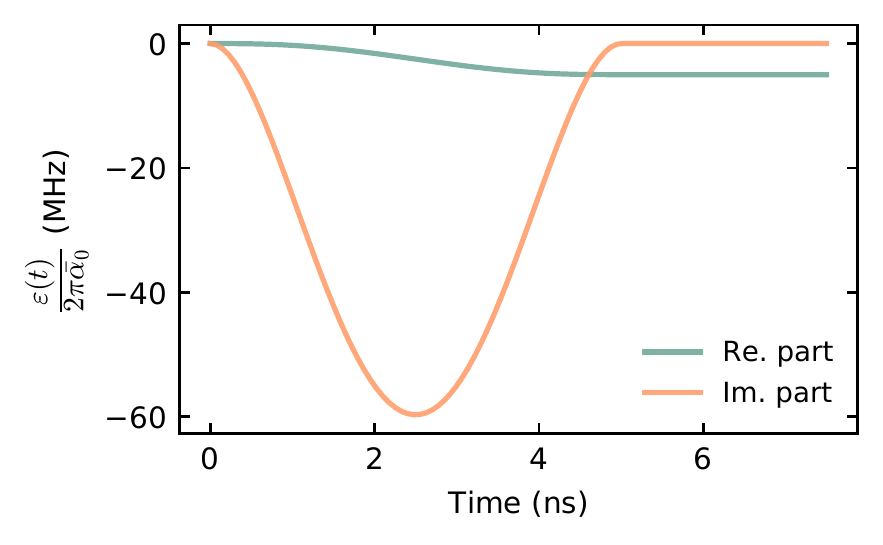}
    \caption{Envelope~$\varepsilon(t)$ in~\cref{eq:enveloppe} used for~\cref{fig:Rabi} to turn off the coupler. The drive can be turned off after some arbitrary time in order to turn the coupler back on with the time-reversed pulse shape shown here [c.f. \cref{app:TQD}].
    }
    \label{fig:tqd_scheme}
\end{figure}

The envelope in~\cref{eq:enveloppe} used in~\cref{fig:Rabi} is shown in~\cref{fig:tqd_scheme}. The ramping time~$\tau$ was set to 5 ns independently of~$\bar{\alpha}_0$. 

\subsection{Inverse participation ratio (IPR) \label{app:IPR}}

\begin{figure}[t!]
    \centering
    \includegraphics[width=0.5\textwidth]{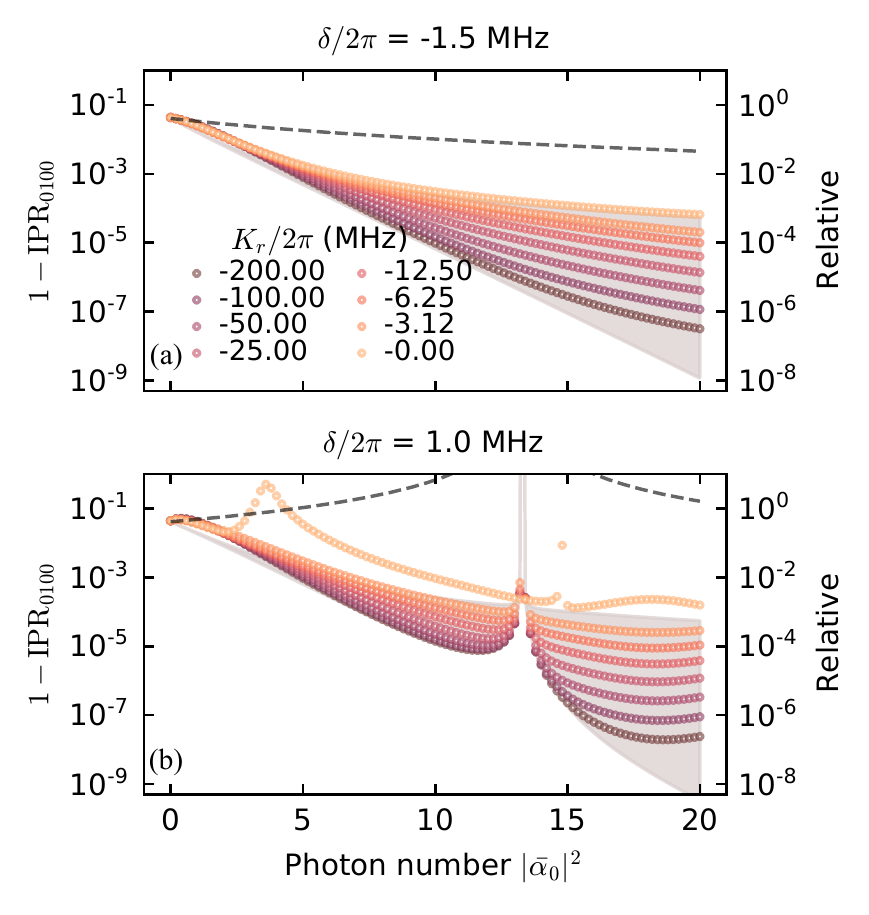}
    \caption{Suppression of~$1-IPR_{0100}$ with respect to the NLR photon number~$\abs{\bar{\alpha}_0}^2$ in the stabilized ground state. Each data point is computed from numerical diagonalization with a fixed drive amplitude in the NLR. Black lines correspond to the same system but with no drive on the NLR and the bus frequency tuned to the ac-Stark shifted frequency found in the driven system. The gray regions are bounded by the analytical estimates in~\cref{eq:ipr0100_analytical} for~$K_r\rightarrow \infty$ and~$K_r = 0$. We also note the presence of a resonance in b) for~$K_r=0$ only for a small photon number for this specific choice of system parameters. This results from frequency collisions with higher Fock states in the NLR which can be otherwise prevented by the addition of a NLR anharmonicity. Here~$(\omega_{1}-\omega_b)/2\pi = 7.0$ MHz,~$(\omega_{2}-\omega_b)/2\pi = 14.0$ MHz,~$K_{1}/2\pi = K_{2}/2\pi = -300.0$ MHz,~$\chi/2\pi = -20.0$ MHz, and~$g/2\pi = 2.0$ MHz.}
    \label{fig:IPR_2}
\end{figure}

In analogy to~\cref{fig:IPR}, the inverse participation ratio for the second qubit, IPR$_{0100}$, is computed numerically and reported in~\cref{fig:IPR_2}. The observations made in~\cref{fig:IPR} can be extended to~\cref{fig:IPR_2}. The only difference here is the emergence of a second divergence at small photon numbers for~$K_r =0$. This peak results from the frequency collisions with higher energy levels in the NLR and the specific choice of parameters. This effect is however absent in the presence of anharmonicity in the NLR. \\

We now provide an analytical estimate for the inverse participation ratio based on a Schrieffer-Wolff (SW) transformation on~\cref{eq:Hp}, where the Hamiltonian takes the form  
\begin{align}
    \hat{H}_I = e^{i\int dt \left(\hat{H}^P-\hat{H}_\kappa^P\right)}\ \hat{H}_g^P \ e^{-i\int dt \left(\hat{H}^P-\hat{H}_\kappa^P\right)}. \label{eq:HI}
\end{align}
To this end we define the generator 
\begin{align}
    \hat{S}_I = i\int dt \ \hat{H}_I, \label{eq:SI}
\end{align}
under which the Hamiltonian transforms to~$\hat{H}_S = e^{\hat{S}_I} \hat{H}_I e^{-\hat{S}_I} + i \dot{\hat{S}}_I = \mathcal{O}(g_j^2)$. In what follows we describe different cases where~$\hat H_g^P$ and~$\hat{H}^P-\hat{H}_\kappa^P$ have a particular time dependence used for parametric modulations in~\cref{app:DD}. Furthermore, we compute the generator back in the polaron frame as 
\begin{align}
    \hat{S} = e^{-i\int dt \left(\hat{H}^P-\hat{H}_\kappa^P\right)}\ \hat{S}_I \ e^{i\int dt \left(\hat{H}^P-\hat{H}_\kappa^P\right)}. \label{eq:S}
\end{align}
This transformation holds for~$||\hat S|| \ll 1$, i.e.~if the transition amplitudes are much smaller in magnitude than the energy gaps. Then, the hybridized states are approximately given by~$\ket{\psi_{h,\nu}} = e^{\hat{S}} \ket{\psi_{b,\nu}} \approx  [1 + \hat{S} + \mathcal{O}(g_j^2)]\ket{\psi_{b,\nu}}$, with~$\ket{\psi_{b,\nu}}$ the bare eigenstates of the full system for~$g_j=0$. An estimation of~\cref{eq:IPR} using the SW transformation follows as
\begin{align}
    \mathrm{IPR}_{\mu} \approx 1 - 2 \matrixel{\psi_{b,\mu}}{\hat{S}^{\dagger}\hat{S}}{\psi_{b,\mu}}.\label{eq:IPR_SW}
\end{align} 
However, as~$\hat{S}$ can be time-dependent, it is useful to also define the time-averaged quantity
\begin{align}
    & \left<\mathrm{IPR}_{\mu}\right>_t \approx 1 - 2 \matrixel{\psi_{b,\mu}}{\hat{S}^{\dagger}\hat{S}}{\psi_{b,\mu}}_t = 1 - 2 \lim_{t\rightarrow \infty}\frac{1}{t} \int dt \ \matrixel{\psi_{b,\mu}}{\hat{S}^{\dagger}\hat{S}}{\psi_{b,\mu}}.\label{eq:IPR_def}
\end{align} 
We will now identify the generator of the SW transformation for the different cases considered in this manuscript.

\subsubsection{Static~$\hat H_g^P$ and~$\hat{H}^P-\hat{H}_\kappa^P$}\label{app:static_SW}

The generator of the SW transformation in this case is 
\begin{align}
    & \hat{S} = \sum_{j,n,m,k,\ell} \frac{\sqrt{(n+1)(m+1)}\matrixel{k}{\hat{D}_{n,r}}{\ell} g_je^{i\phi_n}/\tilde{\Delta}_j}{1 + (m K_j-n \tilde{K}_b)/\tilde{\Delta}_j +q_{j,n,k}-q_{j,n+1,\ell}} \ketbra{m+1}{m}_{j}\ketbra{n}{n+1}_b\ketbra{k}{\ell}_r-\mathrm{h.c.}, \label{eq:generator}\\
    & q_{j,n,k} = \frac{\delta + n \chi  }{\tilde{\Delta}_j}k+\frac{K_r}{\tilde{\Delta}_j} \frac{k (k-1)}{2}.
\end{align}
Given that~$|\langle k|\hat{D}_{n,r}|\ell\rangle|$ is more strongly suppressed for small Fock state numbers~$\left\{k, \ \ell\right\}$, it is clear that the parameters~$(\delta+n\chi)/\tilde{\Delta}_j$ and~$K_r/\tilde{\Delta}_j$, which control the probability of virtually populating larger Fock states of the NLR, play an important role in the efficiency of the suppression of two-qubit interactions. More precisely, we find that
\begin{align}
    \matrixel{\psi_{1000}}{\hat S^{\dagger}\hat{S}}{\psi_{1000}} =e^{-\abs{\bar{\alpha}}^2}\left(\frac{g_j}{\tilde{\Delta}_j}\right)^2\sum_{\ell=0}^{\infty} \frac{\abs{\bar{\alpha}}^{2\ell}/\ell!}{\left(1 -q_{j,1,\ell}\right)^2} = e^{-\abs{\bar{\alpha}}^2} \left(\frac{g_1}{\tilde{\Delta}_1}\right)^2 {}_4 F_4 \left(\boldsymbol{p}_1; 1 + \boldsymbol{p}_1; \abs{\bar{\alpha}}^2\right). \label{eq:IPR_calc}
\end{align}
Throughout this work, we consider the two limiting cases including~$K_r = 0$ and~$K_r \rightarrow \infty$. In these limits, it is possible to derive asymptotic expressions for the inverse participation ratio, as 
\begin{align}
    & \mathrm{IPR}_{1000}^{\abs{K_r}\rightarrow 0} = 1 - 2 \frac{g_1^2}{\tilde{\Delta}_1^2}e^{-\abs{\tilde{\alpha}}^2} {}_2 F_2 \left(\boldsymbol{q}_1; 1 + \boldsymbol{q}_1; \abs{\tilde{\alpha}}^2\right), \label{eq:up}\\
    & \mathrm{IPR}_{1000}^{\abs{K_r}\rightarrow \infty} = 1 - 2 \frac{g_1^2}{\tilde{\Delta}_1^2}e^{-\abs{\tilde{\alpha}}^2} \left(1 + \frac{\abs{\tilde{\alpha}}^2}{\left(1 + \zeta_j\right)^2}\right) 
    \label{eq:down},
\end{align}
where~$\boldsymbol{q}_j = \begin{pmatrix} 1/\zeta_j & 1/\zeta_j \end{pmatrix}$ and~$\zeta_j = -\left(\delta+\chi\right)/\tilde{\Delta}_j$. This can be easily generalized to the states~$0100$ and~$1100$.

\subsubsection{Time-dependent~$\hat H_g^P$ and static~$\hat{H}^P-\hat{H}_\kappa^P$}\label{app:PAM_SW}

We now consider a time-dependent phase
\begin{align}
    e^{i\phi_n} = \sum_{s=-\infty}^{+\infty} \zeta_{n,s} e^{i s \omega_m t},
\end{align}
where~$\zeta_s$ and~$\omega_m$ are free time-independent parameters. In this case, the generator takes the form
\begin{align}
    & \hat{S} = \sum_{j,n,m,k,\ell,s} \frac{\sqrt{(n+1)(m+1)}\matrixel{k}{\hat{D}_{n,r}}{\ell} g_j \zeta_{n,s} e^{i s \omega_m t}/\tilde{\Delta}_j}{1 + s \omega_m/\tilde{\Delta}_j + (m K_j-n \tilde{K}_b)/\tilde{\Delta}_j +q_{j,n,k}-q_{j,n+1,\ell}} \ketbra{m+1}{m}_{j}\ketbra{n}{n+1}_b\ketbra{k}{\ell}_r-\mathrm{h.c.}.
\end{align}
It then follows that 
\begin{align}
    \matrixel{\psi_{1000}}{\hat S^{\dagger}\hat{S}}{\psi_{1000}} =e^{-\abs{\bar{\alpha}}^2}\left(\frac{g_j}{\tilde{\Delta}_j}\right)^2\sum_{\ell,s_1,s_2} \frac{\zeta_{0,s_1}\zeta_{0,s_2}^* e^{i(s_1-s_2)\omega_m t}\abs{\bar{\alpha}}^{2\ell}/\ell!}{\left(1 + s_1 \omega_m/\tilde{\Delta}_j -q_{j,1,\ell}\right)\left(1 + s_2 \omega_m/\tilde{\Delta}_j -q_{j,1,\ell}\right)},
\end{align}
with time average
\begin{align}
    \matrixel{\psi_{1000}}{\hat S^{\dagger}\hat{S}}{\psi_{1000}}_t =e^{-\abs{\bar{\alpha}}^2}\left(\frac{g_j}{\tilde{\Delta}_j}\right)^2\sum_{\ell,s} \frac{\abs{\zeta_{0,s}}^2 \abs{\bar{\alpha}}^{2\ell}/\ell!}{\left(1 + s \omega_m/\tilde{\Delta}_j -q_{j,1,\ell}\right)^2}.
\end{align}

\subsubsection{Static~$\hat H_g^P$ and time-dependent~$\hat{H}^P-\hat{H}_\kappa^P$}\label{app:LD_SW}

Here we consider a time-dependent~$\delta \rightarrow \delta - z \omega_m \sin(\omega_m t)$. The generator takes the form
\begin{align}
    \nonumber \hat{S} = \sum_{j,n,m,k,\ell,s}  \frac{\sqrt{(n+1)(m+1)}\matrixel{k}{\hat{D}_{n,r}}{\ell} g_j e^{i\phi_n}\zeta_{k,\ell,s} e^{i s \omega_m t - i (k-\ell)z\cos(\omega_mt)}/\tilde{\Delta}_j}{1 + s \omega_m/\tilde{\Delta}_j + (m K_j-n \tilde{K}_b)/\tilde{\Delta}_j +q_{j,n,k}-q_{j,n+1,\ell}} & \\
    \cdot \ketbra{m+1}{m}_{j}\ketbra{n}{n+1}_b\ketbra{k}{\ell}_r-\mathrm{h.c.} &,\label{eq: S_long}
\end{align}
where the parameters~$\zeta_{k,\ell,s}$ are defined with
\begin{align}
    e^{i (k-\ell) z \cos(\omega_m t) } = \sum_{s=-\infty}^{+\infty} i^sJ_s\left[(k-\ell)z\right]e^{is \omega_m t} = \sum_{s=-\infty}^{+\infty} \zeta_{k,\ell,s}e^{is\omega_m t},
\end{align}
where we have used a Jacobi-Anger expansion. We thus arrive at
\begin{align}
    \matrixel{\psi_{1000}}{\hat S^{\dagger}\hat{S}}{\psi_{1000}} =e^{-\abs{\bar{\alpha}}^2}\left(\frac{g_j}{\tilde{\Delta}_j}\right)^2\sum_{\ell,s_1,s_2} \frac{(-1)^{s_1+s_2}J_{s_1}(\ell z)J_{s_2}(\ell z) i^{s_1-s_2}e^{i(s_1-s_2)\omega_m t}\abs{\bar{\alpha}}^{2\ell}/\ell!}{\left(1 + s_1 \omega_m/\tilde{\Delta}_j -q_{j,1,\ell}\right)\left(1 + s_2 \omega_m/\tilde{\Delta}_j -q_{j,1,\ell}\right)},
\end{align}
and the time-averaged version 
\begin{align}
    \matrixel{\psi_{1000}}{\hat S^{\dagger}\hat{S}}{\psi_{1000}}_t =e^{-\abs{\bar{\alpha}}^2}\left(\frac{g_j}{\tilde{\Delta}_j}\right)^2\sum_{\ell,s} \frac{J_{s}^2(\ell z) \abs{\bar{\alpha}}^{2\ell}/\ell!}{\left(1 + s \omega_m/\tilde{\Delta}_j -q_{j,1,\ell}\right)^2}. \label{eq:SS_long}
\end{align}

\subsubsection{Time-dependent~$\hat H_g^P$ and~$\hat{H}^P-\hat{H}_\kappa^P$}\label{app:PAM_LD_SW}

Finally, we combine the two previous cases, namely we consider a time-dependent phase
\begin{align}
    e^{i\phi_n} = \sum_{s=-\infty}^{+\infty} \zeta_{n,s}^\phi e^{i s \omega_m^\phi t},
\end{align}
and a time-dependent~$\delta \rightarrow \delta - z \omega_m^\delta \sin(\omega_m^\delta t)$ with
\begin{align}
    e^{i (k-\ell) z \cos(\omega_m^\delta t) } = \sum_{s=-\infty}^{+\infty} i^sJ_s\left[(k-\ell)z\right]e^{is \omega_m^\delta t} = \sum_{s=-\infty}^{+\infty} \zeta_{k,\ell,s}^\delta e^{is\omega_m^\delta t}.
\end{align}
The generator takes the form
\begin{align}
    \nonumber \hat{S} = \sum_{j,n,m,k,\ell,s,r}  \frac{\sqrt{(n+1)(m+1)}\matrixel{k}{\hat{D}_{n,r}}{\ell} g_j \zeta_{n,s}^\phi\zeta_{k,\ell,r}^\delta e^{i \left(s \omega_m^\phi+r\omega_m^\delta\right) t- i(k-\ell)z\cos(\omega_m^\delta t)}/\tilde{\Delta}_j}{1 + \left(s \omega_m^\phi + r \omega_m^\delta\right)/\tilde{\Delta}_j + (m K_j-n \tilde{K}_b)/\tilde{\Delta}_j +q_{j,n,k}-q_{j,n+1,\ell}}& \\
    \cdot \ketbra{m+1}{m}_{j}\ketbra{n}{n+1}_b\ketbra{k}{\ell}_r-\mathrm{h.c.}&.
\end{align}
With this we find that
\begin{align}
    \nonumber \matrixel{\psi_{1000}}{\hat S^{\dagger}\hat{S}}{\psi_{1000}} = & e^{-\abs{\bar{\alpha}}^2}\left(\frac{g_j}{\tilde{\Delta}_j}\right)^2 \sum_{\ell,s_1,s_2,r_1,r_2}i^{s_1-s_2+r_1-r_2}e^{i(s_1-s_2)\omega_m^\phi t+i(r_1-r_2)\omega_m^\delta t}\\
    & \cdot \frac{(-1)^{r_1+r_2}\zeta_{0,s_1}\zeta_{0,s_2}^*J_{r_1}(\ell z)J_{r_2}(\ell z) \abs{\bar{\alpha}}^{2\ell}/\ell!}{\left(1 + \left(s_1\omega_m^\phi + r_1 \omega_m^\delta\right)/\tilde{\Delta}_j -q_{j,1,\ell}\right)\left(1 + \left(s_2\omega_m^\phi + r_2 \omega_m^\delta\right)/\tilde{\Delta}_j -q_{j,1,\ell}\right)},
\end{align}
which under time-averaging reduces to
\begin{align}
    \matrixel{\psi_{1000}}{\hat S^{\dagger}\hat{S}}{\psi_{1000}}_t =e^{-\abs{\bar{\alpha}}^2}\left(\frac{g_j}{\tilde{\Delta}_j}\right)^2\sum_{\ell,s,r} \frac{\abs{\zeta_{0,s}}^2J_{r}^2(\ell z) \abs{\bar{\alpha}}^{2\ell}/\ell!}{\left(1 + \left(s\omega_m^\phi+r \omega_m^\delta\right)/\tilde{\Delta}_j -q_{j,1,\ell}\right)^2}. 
\end{align}
Here we assumed~$s \omega_m^\phi + r \omega_m^\delta =0$ only for~$s=0$ and~$r=0$.

\subsection{Two-qubit interactions\label{app:qubit_qubit_interactions}}

In this section we demonstrate how the matrix elements of the displacement operator in~$\hat{H}_g^P$ [c.f.~\cref{eq:Hp}] yields exponentially suppressed two-qubit interactions by deriving an upper bound based on the inverse participation ratio. Consider the time-evolution operator in the interaction picture
\begin{align}
    \hat{U}_I(t) = \mathcal T e^{-i\int_{0}^{t} d \tau \hat H_I(\tau)} = \sum_{n=0}^\infty \frac{(-i)^n}{n!} \int_{0}^{t} d\tau_1 \cdots \int_{0}^{t} d\tau_n \mathcal{T} \hat H_I(\tau_1) \cdots \hat H_I(\tau_n), \label{eq:UIt}
\end{align}
where~$\mathcal{T}$ is the time-ordering operator and we defined the interaction picture Hamiltonian in~\cref{eq:HI}. It is convenient to approximate~\cref{eq:UIt} using a Magnus expansion,
\begin{align}
    \hat{U}_I = e^{-i\int_0^t d\tau \hat{H}^M_I(\tau)},    
\end{align} 
where~$\hat{H}_I^M$ is an effective Hamiltonian. Up to fourth order in~$g_j$ we have that
\begin{align}
    \nonumber \hat H_I^M(t) \approx & \hat{H}_I(t) - \frac{1}{2}\comm{\hat{H}_I(t)}{\hat{S}_I(t)} + \frac{i}{6}\int_0^t d\tau_1 \left(\comm{\hat{H}_I(t)}{\comm{\hat{H}_I(\tau_1)}{\hat{S}_I(\tau_1)}}+\comm{\hat{S}_I(\tau_1)}{\comm{\hat{H}_I(\tau_1)}{\hat{H}_I(t)}}\right) \\
    \nonumber & + \frac{1}{12} \int_0^t \int_0^{\tau_1} d\tau_1d\tau_2 \left(\comm{\comm{\comm{\hat{H}_I(t)}{\hat{H}_I(\tau_1)}}{\hat{H}_I(\tau_2)}}{\hat{S}_I(\tau_2)} + \comm{\hat H_I(t)}{\comm{\comm{\hat{H}_I(\tau_1)}{\hat{H}_I(\tau_2)}}{\hat{S}_I(\tau_2)}}\right) \\
    & + \frac{1}{12} \int_0^t \int_0^{\tau_1}d\tau_1d\tau_2 \left(\comm{\hat{H}_I(t)}{\comm{\hat{H}_I(\tau_1)}{\comm{\hat{H}_I(\tau_2)}{\hat{S}_I(\tau_2)}}}+\comm{\hat{H}_I(\tau_1)}{\comm{\hat{H}_I(\tau_2)}{\comm{\hat{S}_I(\tau_2)}{\hat{H}_I(t)}}}\right), \label{eq:HIM}
\end{align}
where based on~\cref{eq:SI} we introduced~$\hat{S}_I(t) = i\int_0^t d\tau \hat{H}_I(\tau)$. 

To compute effective two-qubit interactions, we project~$\hat{H}_I^M$ in the ground state of the coupler, valid in the dispersive coupling limit. Due to the form of~$\hat{H}_I$ [c.f.~\cref{eq:HI}], only terms of even order in~$g_j$ remain in~$\hat{H}_I^M$. Defining the projection operator~$\hat{P} = \ketbra{0_b,0_r}{0_b,0_r}$, we compute the effective two-qubit Hamiltonian
\begin{align}
    \hat{H}_I^{q-q}(t) = Tr_c\left(\hat{H}_I^M(t) \hat{P}\right), \label{eq:HIqq}
\end{align}
corresponding to~$\hat{H}_I^M$, where the coupler mode is traced out assuming that it is stabilized in its ground state. For compactness, we define~$\hat{H}_I^M = \sum_{n=1}^4 \hat{H}_{I,n}^M$ where~$\hat{H}_{I,n}^M$ incorporates the couplings~$g_j$ at $n$th order. In what follows we shall approximate the qubits and coupler as two-level systems. This approximation is valid in the limit where the qubit-coupler detunings are small compared to the anharmonicities of the qubits and the coupler. Our results can, however, be extended to include the effects of finite anharmonicity. We therefore write
\begin{align}
    & \hat{H}_I(t) \approx \sum_{j,k,\ell} g_j e^{i\tilde{\Delta}_j\left(1+q_{j,0,k}-q_{j,1,\ell}\right)t}\hat{A}_{k,\ell}\hat{\sigma}_{+,j} + \text{h.c.}, \\
    & \hat{A}_{k,\ell} =  e^{i\phi_0 }\matrixel{k}{\hat{D}_{0,r}}{\ell} \ketbra{k}{\ell}_r \hat{\sigma}_{-,b} , \\
    & q_{j,n,k} = \frac{\delta+n\chi}{\tilde{\Delta}_j} k + \frac{K_r}{\tilde{\Delta}_j}\frac{k(k-1)}{2} = \frac{r_{n,k}}{\tilde{\Delta}_j},
\end{align}
where~$\hat{\sigma}_{-,j}$ ($\hat{\sigma}_{-,b}$) are the spin ladder operators, the two-level approximation of~$\hat{q}_j$ ($\hat{b}$). The interaction picture generator~$\hat{S}_I$ and the polaron frame generator~$\hat{S}$ then take the form
\begin{align}
    & \hat{S}_I(t) \approx \sum_{j,k,\ell} \frac{g_j }{\tilde{\Delta}_j}e^{i\tilde{\Delta}_j\left(1+q_{j,0,k}-q_{j,1,\ell}\right)t}\frac{\hat{A}_{k,\ell}\hat{\sigma}_{+,j}}{1+q_{j,0,k}-q_{j,1,\ell}} - \text{h.c.}, \\
    & \hat{S} \approx \sum_{j,k,\ell} \frac{g_j }{\tilde{\Delta}_j}\frac{\hat{A}_{k,\ell}\hat{\sigma}_{+,j}}{1+q_{j,0,k}-q_{j,1,\ell}} - \text{h.c.},
\end{align}
in analogy to~\cref{eq:SI} and~\cref{eq:S}. Our goal is to find an upper bound on the amplitude of two-qubit interactions in~\cref{eq:HIqq} corresponding to a partial trace over~\cref{eq:HIM}. This upper bound can be obtained from the Cauchy-Schwartz inequality,
\begin{align}
    \abs{Tr\left(\hat{A}\hat{B}\right)} \leq \sqrt{Tr\left(\hat{A}^\dagger\hat{A}\right)} \sqrt{Tr\left(\hat{B}^\dagger\hat{B}\right)} = \abs{\abs{\hat{A}}}_F \abs{\abs{\hat{B}}}_F, \label{eq:CSW}
\end{align}
where~$\abs{\abs{\bullet}}_F$ stands for the Frobenius norm. We will separate the two-qubit interactions by the order in the coupling strengths~$g_j$ and stop at fourth order. However, the following analysis can be extended to higher orders in the coupling strengths.

\subsubsection{Second-order interactions}

We now focus on the two-qubit interactions of second order in the coupling strengths in~\cref{eq:HIqq}. To this end, it is convenient to expand
\begin{align}
    Tr\left(\hat{H}_{I,2}^M(t)\hat{P}\right) =  -\frac{1}{2}Tr_c\left(\hat{H}_I(t)\hat{S}_I(t)\hat{P}\right) 
    + \text{h.c.}, \label{eq:TrHI2}
\end{align}
where
\begin{align}
    Tr_c\left(\hat{H}_I(t)\hat{S}_I(t)\hat{P}\right) = \sum_{s_1,s_2=1}^2 e^{i\left(\tilde{\Delta}_{q_{s_1}}-\tilde{\Delta}_{q_{s_2}}\right)t}g_{s_1} Tr\left(\sum_{\ell=0}^{\infty}\hat{A}_{0,\ell}\frac{g_{s_2}}{\tilde{\Delta}_{q_{s_2}}}\frac{\hat{A}_{0,\ell}^\dagger}{1-q_{s_2,1,\ell}}\hat{P}\right)\hat{\sigma}_{+,s_1}\hat{\sigma}_{-,s_2} .
\end{align}
Interestingly, applying~\cref{eq:CSW} yields
\begin{align}
   \abs{Tr\left(\sum_{\ell=0}^\infty \hat{A}_{0,\ell} \frac{g_{s_2}}{\tilde{\Delta}_{q_{s_2}}}\frac{\hat{A}_{0,\ell}^\dagger}{1-q_{s_2,1,\ell}}\hat{P}\right)} \leq \abs{\abs{\sum_{\ell=0}^\infty\hat{A}_{0,\ell}}}_F \abs{\abs{\sum_{\ell=0}^\infty\frac{g_{s_2}}{\tilde{\Delta}_{q_{s_2}}}\frac{\hat{A}_{0,\ell}^\dagger}{1-q_{s_2,1,\ell}}\hat{P}}}_F&.
\end{align}
where 
\begin{align}
    \abs{\abs{\sum_{\ell=0}^\infty\hat{A}_{0,\ell}}}_F = \abs{\abs{\ketbra{0}{0}_r e^{i\phi_0 } \hat{D}_{0,r}}}_F\cdot\abs{\abs{\hat{\sigma}_{-,b}}}_F = 1.
\end{align}
Furthermore, 
\begin{align}
    \abs{\abs{\sum_{\ell=0}^\infty\frac{g_{s_2}}{\tilde{\Delta}_{q_{s_2}}}\frac{\hat{A}_{0,\ell}^\dagger}{1-q_{s_2,1,\ell}}\hat{P}}}_F & = \sqrt{\left(\frac{g_{s_2}}{\tilde{\Delta}_{q_{s_2}}}\right)^2\sum_{\ell=0}^\infty\frac{\abs{\matrixel{\ell}{\hat{D}_{0,r}}{0}}^2 }{\left(1-q_{s_2,1,\ell}\right)^2}} \approx  \sqrt{\frac{1-\mathrm{IPR}_{\Omega(s_2)}}{2}},
\end{align}
with~$\Omega(s_2) = 1000$ for~$s_2=1$ and~$0100$ for~$s_2=2$, and where we used the dispersive limit expression for~$\mathrm{IPR}$ [c.f.~\cref{eq:IPR_calc}]. Since each term in~\cref{eq:TrHI2} is bounded in magnitude by~$\text{max}_{\nu \in \{1000,0100\}}\sqrt{1-\mathrm{IPR}_\nu}$ it follows that~\cref{eq:TrHI2} is equally bounded in magnitude by~$\text{max}_{\nu \in \{1000,0100\}}\sqrt{1-\mathrm{IPR}_\nu}$. Intuitively, this means that the virtual interaction is always smaller in magnitude than~$\abs{g_ig_j/\tilde{\Delta}_j}$.

\subsubsection{Fourth-order interactions}

We will now demonstrate that the~$\text{max}_{\nu \in \{1000,0100\}}\sqrt{1-\mathrm{IPR}_\nu}$ bound holds true for higher-order two-qubit interactions. First, we expand the commutators such that 
\begin{align}
    \nonumber Tr_c\left(\hat{H}_{I,4}^M\hat{P}\right) & =  \frac{1}{12} \int_0^t \int_0^{\tau_1}  d\tau_1d\tau_2 \left[ 2Tr_c\left(\hat{H}_I(t)\hat{H}_I(\tau_1)\hat{H}_I(\tau_2)\hat{S}_I(\tau_2)\hat{P}\right) \right. \\
    \nonumber & - Tr_c\left(\comm{\hat{H}_I(\tau_1)}{\hat{H}_I(t)}\hat{H}_I(\tau_2)\hat{S}_I(\tau_2)\hat{P}\right) - Tr_c\left(\comm{\hat{H}_I(\tau_2)}{\hat{H}_I(t)}\hat{H}_I(\tau_1)\hat{S}_I(\tau_2)\hat{P}\right) \\
    \nonumber & - Tr_c\left(\comm{\hat{H}_I(\tau_1)}{\hat{H}_I(\tau_2)}\hat{H}_I(t)\hat{S}_I(\tau_2)\hat{P}\right) - Tr_c\left(\comm{\hat{H}_I(t)}{\hat{H}_I(\tau_1)}\hat{S}_I(\tau_2)\hat{H}_I(\tau_2)\hat{P}\right) \\
    &  - Tr_c\left(\comm{\hat{H}_I(t)}{\hat{H}_I(\tau_2)}\hat{S}_I(\tau_2)\hat{H}_I(\tau_1)\hat{P}\right)  \left. - Tr_c\left(\comm{\hat{H}_I(\tau_1)}{\hat{H}_I(\tau_2)}\hat{S}_I(\tau_2)\hat{H}_I(t)\hat{P}\right)\right] + \text{h.c.}.
\end{align}
Second, we find that 
\begin{align}
   \nonumber & Tr_c\left( \hat{H}_I(t_1)\hat{H}_I(t_2)\hat{H}_I(t_3)\hat{S}_I(t_4) \hat{P}\right) = \sum_{s_1,\cdots,s_4=1}^2 e^{i\tilde{\Delta}_{q_{s_1}}t_1}e^{-i\tilde{\Delta}_{q_{s_2}}t_2}e^{i\tilde{\Delta}_{q_{s_3}}t_3} e^{-i\tilde{\Delta}_{q_{s_4}}t_4}g_{s_1}g_{s_2}g_{s_3} \\
   & \cdot  Tr\left(\sum_{k,\ell,m=0}^{\infty}e^{-i r_{1,k}(t_1-t_2)-ir_{0,\ell}(t_2-t_3)-ir_{1,m}(t_3-t_4)}\hat{A}_{0,k}\hat{A}_{\ell,k}^\dagger\hat{A}_{\ell,m}\frac{g_{s_4}}{\tilde{\Delta}_{q_{s_4}}}\frac{\hat{A}_{0,m}^\dagger}{1-q_{s_4,1,m}}\hat{P}\right)\hat{\sigma}_{+,s_1}\hat{\sigma}_{-,s_2} \hat{\sigma}_{+,s_3}\hat{\sigma}_{-,s_4} , \label{eq:Tr1}
\end{align}
and similarly,
\begin{align}
   \nonumber & Tr_c\left( \hat{H}_I(t_1)\hat{H}_I(t_2)\hat{S}_I(t_3)\hat{H}_I(t_4) \hat{P}\right) = -\sum_{s_1,\cdots,s_4=1}^2 e^{i\tilde{\Delta}_{q_{s_1}}t_1}e^{-i\tilde{\Delta}_{q_{s_2}}t_2}e^{i\tilde{\Delta}_{q_{s_3}}t_3} e^{-i\tilde{\Delta}_{q_{s_4}}t_4}g_{s_1}g_{s_2}g_{s_4} \\
   & \cdot  Tr\left(\sum_{k,\ell,m=0}^{\infty}e^{-i r_{1,k}(t_1-t_2)-ir_{0,\ell}(t_2-t_3)-ir_{1,m}(t_3-t_4)}\hat{A}_{0,k}\hat{A}_{\ell,k}^\dagger\frac{g_{s_3}}{\tilde{\Delta}_{q_{s_3}}}\frac{\hat{A}_{\ell,m}\hat{A}_{0,m}^\dagger}{1+q_{s_3,0,\ell}-q_{s_3,1,m}}\hat{P}\right)\hat{\sigma}_{+,s_1}\hat{\sigma}_{-,s_2} \hat{\sigma}_{+,s_3}\hat{\sigma}_{-,s_4}.\label{eq:Tr2}
\end{align}
As it was done in the previous section we apply~\cref{eq:CSW} to find an upper bound on~\cref{eq:Tr1}
\begin{align}
   \nonumber & \abs{Tr\left(\sum_{k,\ell,m=0}^{\infty}e^{-i r_{1,k}(t_1-t_2)-ir_{0,\ell}(t_2-t_3)-ir_{1,m}(t_3-t_4)}\hat{A}_{0,k}\hat{A}_{\ell,k}^\dagger\hat{A}_{\ell,m}\frac{g_{s_4}}{\tilde{\Delta}_{q_{s_4}}}\frac{\hat{A}_{0,m}^\dagger}{1-q_{s_4,1,m}}\hat{P}\right)} \\
   \nonumber & \leq \abs{\abs{\sum_{k,\ell,m=0}^{\infty}e^{-i r_{1,k}(t_1-t_2)-ir_{0,\ell}(t_2-t_3)-ir_{1,m}(t_3-t_4)}\hat{A}_{0,k}\hat{A}_{\ell,k}^\dagger\hat{A}_{\ell,m}}}_F \abs{\abs{\sum_{m=0}^{\infty}\frac{g_{s_4}}{\tilde{\Delta}_{q_{s_4}}}\frac{\hat{A}_{0,m}^\dagger}{1-q_{s_4,1,m}}\hat{P}}}_F \\
   & \leq \abs{\abs{\sum_{k,\ell,m=0}^{\infty}\hat{A}_{0,k}\hat{A}_{\ell,k}^\dagger\hat{A}_{\ell,m}}}_F \abs{\abs{\sum_{m=0}^{\infty}\frac{g_{s_4}}{\tilde{\Delta}_{q_{s_4}}}\frac{\hat{A}_{0,m}^\dagger}{1-q_{s_4,1,m}}\hat{P}}}_F, 
 \end{align}
where it can also be verified that~$\abs{\abs{\sum_{k,\ell,m=0}^{\infty}\hat{A}_{0,k}\hat{A}_{\ell,k}^\dagger\hat{A}_{\ell,m}}}_F = 1$ and where
  \begin{align}
      \abs{\abs{\sum_{m=0}^{\infty}\frac{g_{s_4}}{\tilde{\Delta}_{q_{s_4}}}\frac{\hat{A}_{0,m}^\dagger}{1-q_{s_4,1,m}}\hat{P}}}_F  \approx \sqrt{\frac{1-\mathrm{IPR}_{\Omega(s_4))}}{2}}.
 \end{align}
Similarly,~\cref{eq:CSW} can be applied on~\cref{eq:Tr2}
 \begin{align}
    \nonumber & \abs{Tr\left(\sum_{k,\ell,m=0}^{\infty}e^{-i r_{1,k}(t_1-t_2)-ir_{0,\ell}(t_2-t_3)-ir_{1,m}(t_3-t_4)}\hat{A}_{0,k}\hat{A}_{\ell,k}^\dagger\frac{g_{s_3}}{\tilde{\Delta}_{q_{s_3}}}\frac{\hat{A}_{\ell,m}\hat{A}_{0,m}^\dagger}{1+q_{s_3,0,\ell}-q_{s_3,1,m}}\hat{P}\right)} \\
    \nonumber & \leq \abs{\abs{\sum_{k,\ell=0}^{\infty}e^{-i r_{1,k}(t_1-t_2)-ir_{0,\ell}(t_2-t_3)}\hat{A}_{0,k}\hat{A}_{\ell,k}^\dagger}}_F \abs{\abs{\sum_{\ell,m=0}^\infty e^{-ir_{1,m}(t_3-t_4)}\frac{g_{s_3}}{\tilde{\Delta}_{q_{s_3}}}\frac{\hat{A}_{\ell,m}\hat{A}_{0,m}^\dagger}{1+q_{s_3,0,\ell}-q_{s_3,1,m}}\hat{P}}}_F \\
    & \leq \abs{\abs{\sum_{k,\ell=0}^{\infty}\hat{A}_{0,k}\hat{A}_{\ell,k}^\dagger}}_F \abs{\abs{\sum_{\ell,m=0}^\infty \frac{g_{s_3}}{\tilde{\Delta}_{q_{s_3}}}\frac{\hat{A}_{\ell,m}\hat{A}_{0,m}^\dagger}{1+q_{s_3,0,\ell}-q_{s_3,1,m}}\hat{P}}}_F,
\end{align}
where~$\abs{\abs{\sum_{k,\ell=0}^{\infty}\hat{A}_{0,k}\hat{A}_{\ell,k}^\dagger}}_F = 1$ and where
\begin{align}
    \abs{\abs{\sum_{\ell,m=0}^\infty \frac{g_{s_3}}{\tilde{\Delta}_{q_{s_3}}}\frac{\hat{A}_{\ell,m}\hat{A}_{0,m}^\dagger}{1+q_{s_3,0,\ell}-q_{s_3,1,m}}\hat{P}}}_F = \sqrt{\sum_{\ell,m,m'=0}^\infty\left(\frac{g_{s_3}}{\tilde{\Delta}_{q_{s_3}}}\right)^2 \frac{\matrixel{0}{\hat{D}_{0,r}}{m'}\matrixel{m'}{\hat{D}_{0,r}^\dagger}{\ell}\matrixel{\ell}{\hat{D}_{0,r}}{m}\matrixel{m}{\hat{D}_{0,r}^\dagger}{0}}{\left(1+q_{s_3,0,\ell}-q_{s_3,1,m'}\right)\left(1+q_{s_3,0,\ell}-q_{s_3,1,m}\right)}} . &
\end{align}
Considering this norm to be bounded from above by the limiting case~$\delta =0$ and~$K_r= 0$, we find that  
\begin{align}
    \abs{\abs{\sum_{\ell,m=0}^\infty \frac{g_{s_3}}{\tilde{\Delta}_{q_{s_3}}}\frac{\hat{A}_{\ell,m}\hat{A}_{0,m}^\dagger}{1+q_{s_3,0,\ell}-q_{s_3,1,m}}\hat{P}}}_F & \lessapprox \sqrt{\sum_{\ell,m=0}^\infty\left(\frac{g_{s_3}}{\tilde{\Delta}_{q_{s_3}}}\right)^2 \frac{\abs{\matrixel{m}{\hat{D}_{0,r}}{0}}^2}{\left(1-q_{s_3,1,m}\right)^2}}\approx \sqrt{\frac{1-\mathrm{IPR}_{\Omega(s_3)}}{2}}.
\end{align}
As previously found for second-order interactions, all contributions are bounded in magnitude by~$\text{max}_{\nu \in \{1000,0100\}}\sqrt{1-\mathrm{IPR}_\nu}$. These observations can be straightforwardly extended to higher orders in a similar fashion. 

This implies that exponentially suppressing~$1-\mathrm{IPR}_\nu$ will equally exponentially suppress virtual two-qubit interactions. However, we stress that this only provides an estimate for the order of magnitude.

\subsection{Measurement-induced dephasing\label{app:measurement_dephasing}}

\begin{figure*}[t!]
    \centering
    \includegraphics[width=0.8\textwidth]{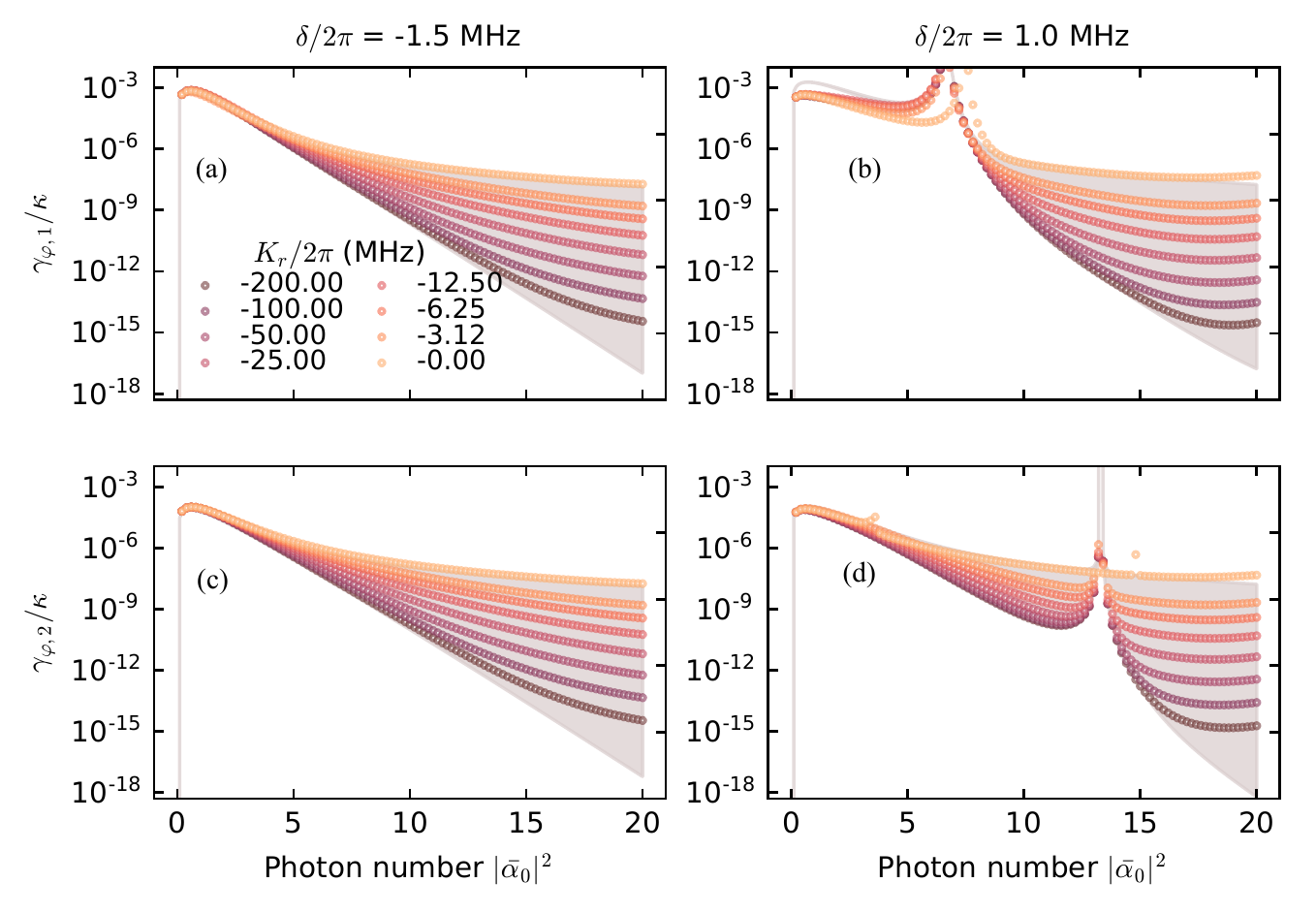}
    \caption{Measurement-induced dephasing rates in the qubits~$1$ and~$2$. Dephasing rate for a)-b) qubit~$1$, and c)-d) qubit~$2$. The gray regions correspond to the analytical estimates in~\cref{eq:dephasing_rates_1} with~\cref{eq:up,eq:down}. Left column (a), c) and e) with~$\delta/2\pi=-1.5$ MHz) and right column (b), d) and f) for~$\delta/2\pi=1.0$ MHz) correspond to two different parameter regimes set by the sign of~$\delta$. Here~$(\omega_{1}-\omega_b)/2\pi = 7.0$ MHz,~$(\omega_{2}-\omega_b)/2\pi = 14.0$ MHz,~$K_{1}/2\pi = K_{2}/2\pi = -300.0$ MHz,~$\chi/2\pi = -20.0$ MHz,~$g/2\pi = 2.0$ MHz, and~$\kappa/2\pi = 100.0$ kHz.}
    \label{fig:dephasing}
\end{figure*}

In order to quantify the dephasing induced by the coupler drive we express~\cref{eq:LbP} in the hybridized eigenbasis
\begin{align}
    & \hat{L}_b^P = \sqrt{\frac{\kappa\abs{\bar{\alpha}}^2}{4}} \sum_{\nu,\nu'}c_{\nu;\nu'} \ketbra{\psi_{h,\nu}}{\psi_{h,\nu'}}, 
\end{align}
with
\begin{equation}
    c_{\nu;\nu'} = \matrixel{\psi_{h,\nu}}{(\ketbra{1}{1}_b-\ketbra{0}{0}_b)}{\psi_{h,\nu'}}.
\end{equation}
To study the effects of bus dephasing on $Q_1$ alone, we first trace out~$Q_2$,~$B$ and~$R$ in the hybridized basis where all three have zero excitation 
\begin{align}
    \hat{L}_1^P \approx \sum_{i,j} \matrixel{\psi_{h,i000}}{\hat{L}_b^P}{\psi_{h,j000}} \ketbra{\psi_{h,i000}}{\psi_{h,j000}}.
\end{align}
We then apply a rotating-wave approximation to obtain 
\begin{align}
    \hat{L}_1^P \approx  \sqrt{\frac{\kappa\abs{\bar{\alpha}}^2}{4}} \sum_{i}c_{i000;i000} \ketbra{\psi_{h,i000}}{\psi_{h,i000}}.
\end{align}
Because the total excitation number in $Q_1$, $Q_2$ and~$B$ is conserved under~\cref{eq:Hp} and the interaction in~\cref{eq:Hp} dominantly yields hybridization between~$1$ and~$B$, we neglect~$\braket{\psi_{h,1000}}{\psi_{b,010k}}\approx 0$. We also use the fact that~$\sum_k \abs{\braket{\psi_{h,0000}}{\psi_{b,000k}}}^2 = 1$ due to total excitation number conservation, and finally,~$\sum_k \abs{\braket{\psi_{h,1000}}{\psi_{b,100k}}}^2 \approx 1 - \sum_k \abs{\braket{\psi_{h,1000}}{\psi_{b,001k}}}^2$. It follows that~$c_{1000;1000} \approx 2 \sum_{k} \abs{\braket{\psi_{h,1000}}{\psi_{b,001k}}}^2 -1$ and~$c_{0000;0000} \approx -1$, leading to
\begin{align}
    \hat{L}_1^P \approx & \sqrt{\frac{\kappa\abs{\bar{\alpha}}^2}{4}} \sum_{k=0}^{\infty} \abs{\braket{\psi_{h,1000}}{\psi_{b,001k}}}^2 \left(\ketbra{\psi_{h,1000}}{\psi_{h,1000}} - \ketbra{\psi_{h,0000}}{\psi_{h,0000}}\right). \label{eq:L1P}
\end{align}
The dephasing rate of $Q_1$ can be estimated by computing the dephasing rate associated with~\cref{eq:L1P}, which takes the form
\begin{equation}
    \begin{split}
        \gamma_{\varphi,1} 
        &= \frac{\kappa \abs{\bar{\alpha}}^2}{2} \left(\sum_{k=0}^{\infty}\abs{\braket{\psi_{h,1000}}{\psi_{b,001k}}}^2\right)^2 \approx \frac{\kappa \abs{\bar{\alpha}}^2}{2}  \frac{1-\mathrm{IPR}_{1000}}{2},
    \end{split}
\end{equation}
where the second line follows from the Schrieffer-Wolff transformation [c.f.~\cref{eq:IPR_SW}]. The expression above was obtained under a rotating-wave approximation, which is valid for~$|\gamma_{\varphi,1}/\tilde{\Delta}_1|\ll 1$. An expression for the second qubit is obtained by simply replacing the subscript~$1000$ by~$0100$. Measurement-induced dephasing rates in the qubits are estimated numerically from diagonalization and are reported in~\cref{fig:dephasing}.

\subsection{Parameter regimes}\label{app:parameter_regimes}

We have seen in~\cref{fig:IPR,fig:ZZ} that the choice of frequency detuning between the NLR and the drive~$\delta$, which controls the sign of the ac-Stark shift between the qubits and the bus, has both a quantitative and qualitative impact on the  system. To better understand this behavior, we consider again~\cref{eq:ipr1000_analytical,eq:ipr0100_analytical} where we now express the generalized hypergeometric functions as
\begin{align}
    {}_4F_4 (\boldsymbol{p}_j,1+\boldsymbol{p}_j;\abs{\bar{\alpha}}^2) = \sum_{n=0}^{\infty}\frac{\abs{\bar{\alpha}}^{2n}}{n!} \prod_{k=\pm}\left(1+ \frac{n}{p_{jk}}\right)^{-2}.
    \label{eq:4F4}
\end{align}
We recall that in this expression~$p_{j\pm} = \beta [1 \pm   \sqrt{ 1  + \smash[b]{2\tilde{\Delta}_j \beta^{-2}/K_r}}]$ with~$\beta = (\delta+\chi)/K_r - 1/2$ and where~$\tilde{\Delta}_j$ is the ac-Stark shifted detuning given in~\cref{eq:DeltaTilde}.
Physically,~\cref{eq:4F4} corresponds to a weighted sum over all virtual transitions to higher energy levels in the NLR for a displacement~$\bar{\alpha}$ during a~$0\rightarrow 1$ transition in the bus. The poles~$1+n/p_{jk}$ in~\cref{eq:4F4} correspond to frequency collisions with these higher energy levels. 

Given that we wish to maximize the exponential suppression of IPR [i.e.~the exponential factor in~\cref{eq:ipr1000_analytical,eq:ipr0100_analytical}], we want~${}_4F_4(\boldsymbol{p}_j,1+\boldsymbol{p}_j;\abs{\bar{\alpha}}^2)$ to be as close as possible to unity. To have a monotonic suppression with respect to~$\abs{\bar{\alpha}}^2$, this parameter should be chosen such as to avoid frequency collisions corresponding to the poles of~\cref{eq:4F4}. With these constraints, we define two key parameter regimes below. In what follows we take $(\kappa/2)^2 \ll \abs{\delta(\delta+\chi)}$, but we note that~$\kappa$ also plays a role in controlling the ac-Stark shift of the bus and can be seen as an additional knob. To simplify the analysis below, it is useful to note that for~$K_r = 0$ we find~$p_{j+} \rightarrow \infty$ and~$p_{j-}\rightarrow -\tilde{\Delta}_j/(\delta+\chi)$, while for~$\abs{K_r}\gg \{|\tilde{\Delta}_j|, \ |\delta+\chi|\}$ we find~$p_{j+} \rightarrow -1 + 2 (\delta+\chi-\tilde{\Delta}_j)/K_r$ and~$p_{j-}\rightarrow 2\tilde{\Delta}_j/K_r$.

\subsubsection{Monotonic suppression}

We first focus on the situation illustrated in panels (a) of~\cref{fig:IPR,fig:ZZ} where there are no frequency collisions between the different modes of the system and the suppression factor is a smooth function of the photon number. There are two possible sources of frequency collisions: i) the ac-Stark shifted qubit-bus detuning~$\tilde{\Delta}_j$ of~\cref{eq:DeltaTilde} and ii) higher energy levels in the NLR as captured by \cref{eq:4F4}. 

First, to avoid a collision where~$\tilde{\Delta}_j =0$, the ac-Stark-shifted qubit-bus detuning should ideally grow in magnitude with respect to~$\bar{\alpha}$. This, in turn implies that~$\omega_{qj}-\omega_b$ and~$\delta(\delta+\chi)/\chi \approx \delta$ should have opposite signs. 
Considering the second source of frequency collisions,~$p_{j\pm}>0$ in~\cref{eq:4F4} suppresses frequency collisions with higher energy levels in the NLR. For~$K_r = 0$ this condition is achieved with~$p_{j-} = -\tilde{\Delta}_j/(\delta+\chi)>0$, meaning that~$\tilde{\Delta}_j$ and~$\delta+\chi$ have opposite signs. Combining this with the above finding, it follows that~$\delta$ and~$\chi$ have the same sign, and opposite sign to~$\omega_{qj}-\omega_b$. We stress that this choice of parameters is equally compatible with the large~$\abs{K_r}$ limit because $\abs{1/p_{j-}}$ is then large. 

Combining the above with the fact that $\abs{\delta/\chi} \ll 1$ for large conditional displacements, allows us to define the parameter regime for monotonic suppression of the IPR as
\begin{align}\label{eq:SmoothParameterRegime}
    \abs{\delta/\chi} \ll 1, \quad \left(\omega_{j}-\omega_b\right)\delta < 0, \quad \left(\omega_{j}-\omega_b\right)\chi  < 0.
\end{align}
Because the nonlinearity of Josephson junctions is negative, in panels (a) of~\cref{fig:IPR,fig:ZZ} we chose~$\chi < 0$ and~$\omega_{j}-\omega_b>0$.

\subsubsection{Nonmonotonic and strong suppression}

As illustrated in panels (b) of~\cref{fig:IPR,fig:ZZ}, working in a parameter regime where the behavior of the IPR with photon number is nonmonotonic can lead to stronger suppression. A first observation to understanding this effect is that, taking~$p_{j-}\propto \tilde{\Delta}_j$ and~$p_{j-}\rightarrow 0$, forces~\cref{eq:4F4} to be unity. Moreover, we can exploit the fact that symmetric two-qubit interactions, such as the~$ZZ$ interaction, are suppressed when the bus frequency lies between the qubit frequencies. 
As a result, the choice~$\tilde{\Delta}_1 + \tilde{\Delta}_2 = 0$ minimizes the~$ZZ$ interaction in~\cref{eq:chi_AC} in the limit of large anharmonicities with respect to the detunings, i.e.~$|\tilde{\Delta}_j/K_j| \ll 1$ and~$|\tilde{\Delta}_j/\tilde{K}_b| \ll 1$. 

Based on these arguments, we define a second parameter regime according to 
\begin{align}
    \abs{\delta/\chi} \ll 1, \quad \left(\omega_{j}-\omega_b\right)\delta > 0, \quad \left(\omega_{j}-\omega_b\right)\chi  < 0.
\end{align}
The essential difference with respect to~\cref{eq:SmoothParameterRegime} is that~$\delta$ has now changed sign. In this regime, the ac-Stark shift of the bus mode changes the sign of the qubit-bus detunings for some~$\bar{\alpha}$, something that can help suppressing spurious interactions. 

It is worth highlighting that the protocol presented here works best for small qubit-bus detunings~$|\tilde{\Delta}_j|$ relative to~$\abs{\chi}$. This is because the drive on the NLR, close to its resonance frequency, renders the energy gaps between the stabilized states in the NLR smaller in magnitude than in the bare energy spectrum of the NLR. Without anharmonicity in the NLR, these energy gaps are predominantly set by~$\abs{\chi}$. 

Finally, although we have focused on~$\bar{\alpha}_0$ with the TQD protocol, one could have chosen to grow~$\bar{\alpha}_1$ for example, such as to still suppress the dominant~$0\leftrightarrow 1$ transition in the bus. The main reason for the focus on~$\bar{\alpha}_0$ is that, in this case,~$\delta$ is made small but~$\delta+\chi$ is large, thus preventing transitions to higher energy states of the NLR.

\section{Effective parametric modulation \label{app:DD}}

In this section we provide supporting analytical derivations and numerical results for the parametric modulation (PAM) scheme presented in~\cref{sec:DD_scheme} and discuss another decoupling scheme based on a longitudinal drive (LD) in the NLR to mimic the effects of an anharmonicity. We also show that the two schemes can be potentially combined to offer stronger suppression of qubit-qubit interactions.

\subsection{PAM: additional tones in the NLR drive}

\begin{figure}[h!]
    \centering
    \includegraphics[width=0.45\textwidth]{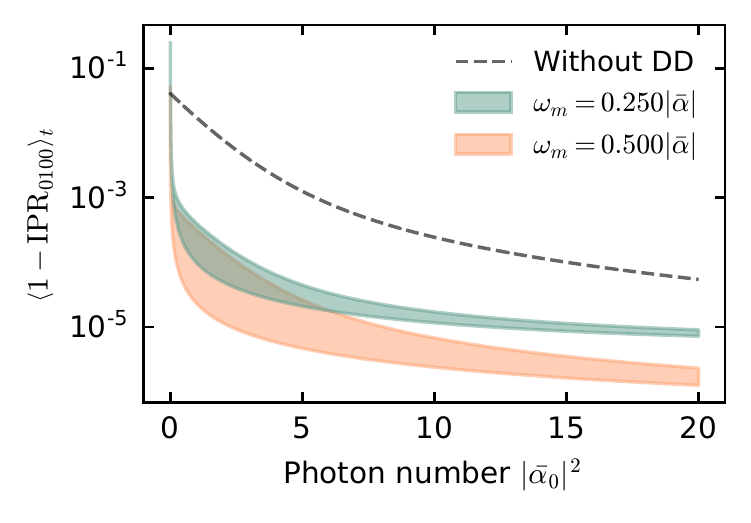}
    \caption{Dynamical decoupling in the bus: time-averaged inverse participation ratio against photon number in the ground state. Here we assume two additional tones~$\omega_r\pm \omega_m$ with amplitude~$\omega_m \lambda /\bar{\alpha}$, and we choose~$\omega_m$ such that the voltage-drive amplitude is fixed for all~$\bar{\alpha}$s. The regions are bounded by~$\lambda = \lambda_0$ where~$J_0(\lambda_0) = 0$ and~$\pm$10\% error on~$\lambda_0$. Note that we exclude anharmonicity in the NLR ($K_r = 0$). Parameters:~$\delta /2\pi=-1.0$ MHz,~$\chi/2\pi = -20.0$ MHz,~$(\omega_{1}-\omega_b)/2\pi = 7.0$ MHz,~$(\omega_{2}-\omega_b)/2\pi = 14.0$ MHz,~$K_{1}/2\pi = K_{2}/2\pi = -300.0$ MHz, and~$g/2\pi = 2.0$ MHz.}\label{fig:IPR_otter_2}
\end{figure}

We comment on the effects of the fast-oscillating contributions in~\cref{eq:Jacobi-Anger}. As previously stated, these terms impose a lower bound on the time-averaged~$1-$IPR. For large~$\omega_m$ as compared to~$\delta$ and~$\chi$, this lower bound is approximately~$2\left(g/\omega_m\right)^2$. In the particular case of~$\omega_m = \omega_0 \abs{\bar{\alpha}}$, such that the voltage drive amplitude in the NLR is displacement-independent, we observe that the asymptotic behavior in the suppression is polynomial in~$\abs{\bar{\alpha}}$. We also stress that an appreciable anharmonicity~$K_r$ in the NLR could in principle result in additional frequency collisions due to negative~$s\omega_m/\tilde{\Delta}_1$ where~$s$ is an integer. Finally and in addition to~\cref{fig:IPR_DD} we report the inverse participation of the second qubit in~\cref{fig:IPR_otter_2}. For both~\cref{fig:IPR_DD,fig:IPR_otter_2} we plot the time-averaged inverse participation ratio obtained in~\cref{app:PAM_SW}.

\subsection{LD: longitudinal drive in the NLR}

\begin{figure}[h!]
    \centering
    \includegraphics[width=0.9\textwidth]{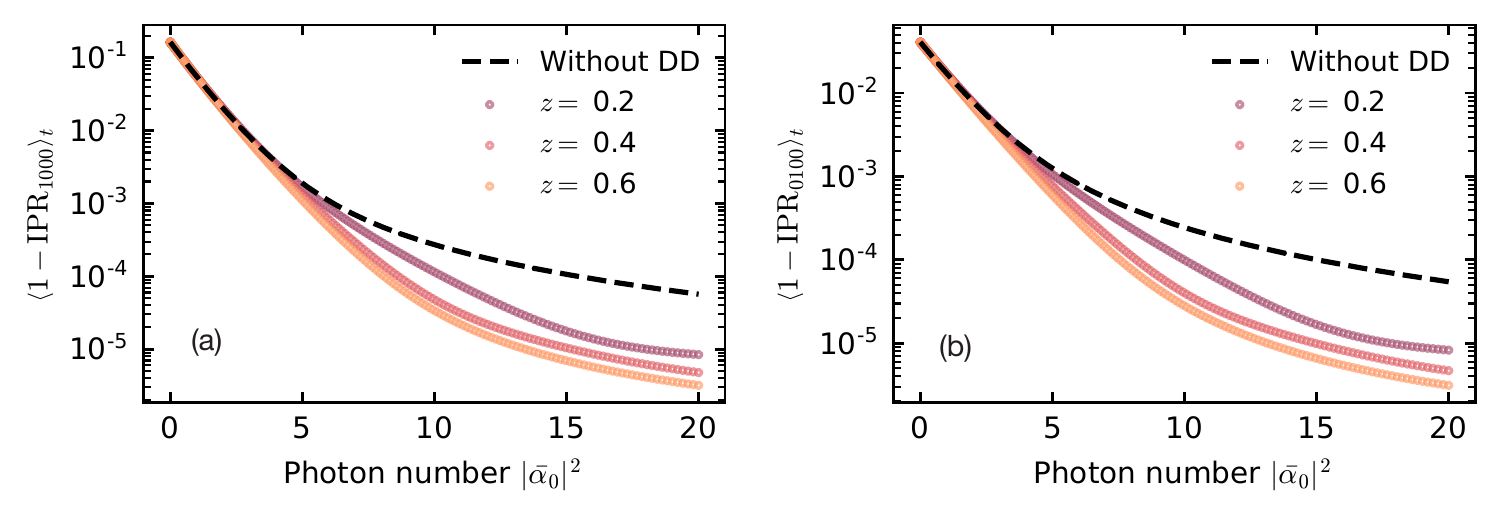}
    \caption{Dynamical decoupling in the NLR: time-averaged inverse participation ratio against photon number in the ground state. We assume an additional longitudinal drive~$\delta\rightarrow \delta - z \omega_m \sin\left(\omega_m t\right)$ in the NLR with~$\omega_m = 1.0$ GHz. Note that no anharmonicity is considered for the NLR ($K_r = 0$). Parameters:~$\delta /2\pi=-1.0$ MHz,~$\chi/2\pi = -20.0$ MHz,~$(\omega_{1}-\omega_b)/2\pi = 7.0$ MHz,~$(\omega_{2}-\omega_b)/2\pi = 14.0$ MHz,~$K_{1}/2\pi = K_{2}/2\pi = -300.0$ MHz, and~$g/2\pi = 2.0$ MHz.\label{fig:IPR_long}}
\end{figure}

We now discuss how a longitudinal drive in the NLR can help recover a strong suppression in absence of large anharmonicity and without the need for fine-tuning. Our starting point is~\cref{eq:SS_long}. In the~$z\rightarrow \infty$ limit, the asymptotic behaviour of the Bessel functions is
\begin{align}
    J_\nu(z) \sim \sqrt{\frac{2}{\pi z}} \cos\left(z - \frac{2\nu + 1}{4}\pi\right),
\end{align}
and we also have~$J_\nu(-z)= (-1)^\nu J_\nu(z)$. It follows that the matrix elements of~\cref{eq: S_long} are renormalized by~$\abs{(k-\ell) z}^{-1/2}$ for~$k\neq \ell$. This result is appealing given that the suppression of two-qubit interactions is strongest if the NLR is constrained to Fock states of small photon number and it is desirable to suppress~$0\rightarrow k$ transitions in the NLR, as shown in~\cref{fig:IPR_long}.

A key advantage of this scheme is that it might be possible to relax parameter constraints for~$\tilde{\Delta}_q$ and~$\chi$ due to the simulated anharmonicity. Regarding the physical implementation, it can range from a modulated detuning in the voltage drive to flux-modulating a superconducting loop with junctions. The results shown in~\cref{fig:IPR_long} correspond to the time-averaged participation ratio obtained in~\cref{app:LD_SW}.

\subsection{PAM and LD in parallel}

\begin{figure}[h!]
    \centering
    \includegraphics[width=0.9\textwidth]{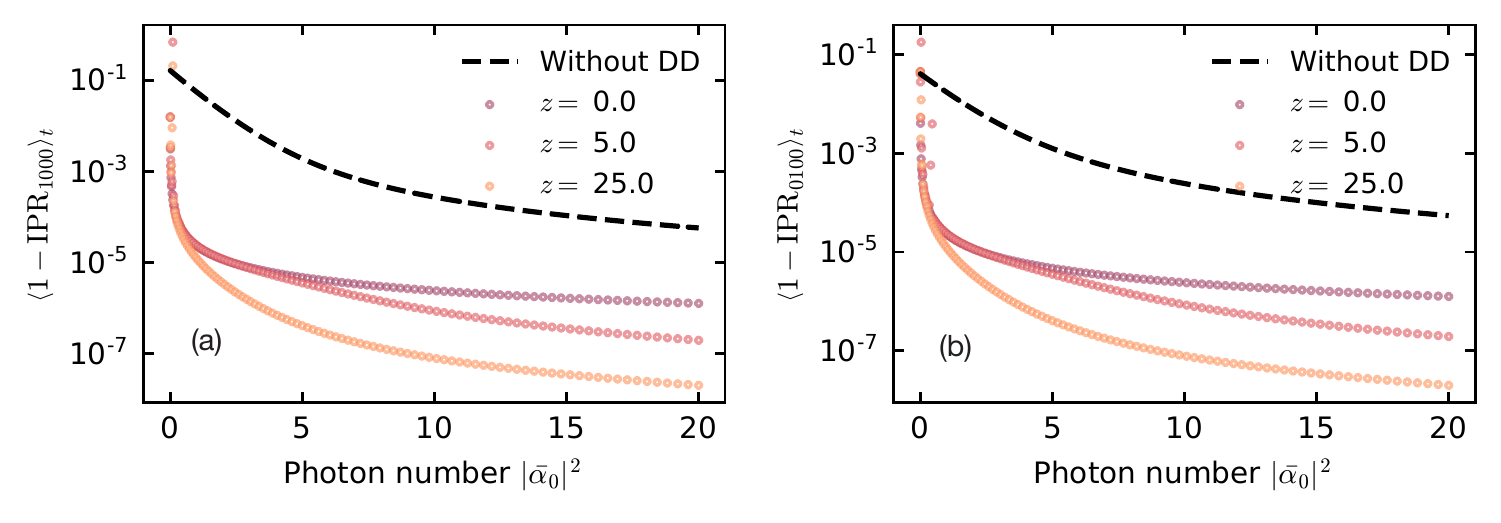}
    \caption{Dynamical decoupling in the bus and in the NLR: time-averaged inverse participation ratio against photon number in the ground state. Here we assume two additional tones~$\omega_r\pm \omega_m^\phi$ in the voltage drive with amplitude~$\omega_m^\phi \lambda /\bar{\alpha}$, and we choose~$\omega_m^\phi = 0.5 \bar{\alpha}$ such that the voltage-drive amplitude is fixed for all~$\bar{\alpha}$s with~$\lambda = \lambda_0$ where~$J_0(\lambda_0) = 0$. We assume an additional longitudinal drive~$\delta\rightarrow \delta - z \omega_m^\delta  \sin\left(\omega_m^\delta t\right)$ in the NLR with~$\omega_m^\delta = 10.0$ MHz. Parameters:~$\delta /2\pi=-1.0$ MHz,~$\chi/2\pi = -20.0$ MHz,~$(\omega_{1}-\omega_b)/2\pi = 7.0$ MHz,~$(\omega_{2}-\omega_b)/2\pi = 14.0$ MHz,~$K_{1}/2\pi = K_{2}/2\pi = -300.0$ MHz, and~$g/2\pi = 2.0$ MHz.\label{fig:IPR_DD_long}}
\end{figure}

A key challenge in combining PAM and LD is to prevent frequency collisions. Having simultaneously large modulation frequencies for both schemes is therefore not a good option. Given that the lower bound on~$1-$IPR is dominantly set by the modulation frequency in PAM, we choose this frequency to be the largest. Since the modulation frequency in LD is small, we can choose~$z$ to be large. We indeed observe in~\cref{fig:IPR_DD_long} that it is possible to grow stronger suppression factors by increasing~$z$.~\cref{fig:IPR_DD_long} illustrates the time-averaged inverse participation obtained in~\cref{app:PAM_LD_SW}.

\section{Superconducting implementation \label{app:circuit}}

\begin{figure}[t!]
    \centering
    \includegraphics[width=0.5\textwidth]{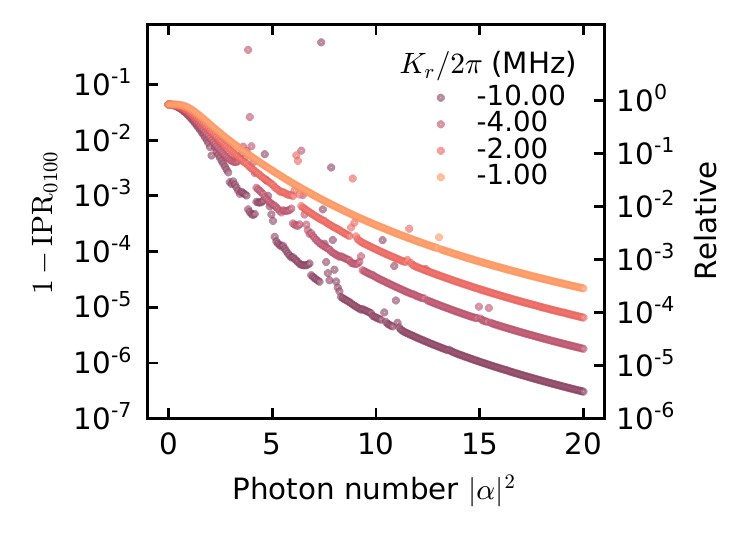}
    \caption{$1-\mathrm{IPR}_{0100}$ estimated with the numerical diagonalization of the full system with the effective Hamiltonian in~\cref{eq:HKerrCat} with~$\delta /2\pi=19.9$ MHz,~$\chi/2\pi = -20.0$ MHz (parameter regime of c) where~$\abs{\alpha_0}\rightarrow 0$). Here~$(\omega_{1}-\omega_b)/2\pi = 7.0$ MHz,~$(\omega_{2}-\omega_b)/2\pi = 14.0$ MHz,~$K_{1}/2\pi = K_{2}/2\pi = -300.0$ MHz, and~$g/2\pi = 2.0$ MHz.\label{fig:circuit_implementation_2}} 
\end{figure}

The inverse participation ratio for the second qubit is illustrated in~\cref{fig:circuit_implementation_2} and should be contrasted to that of panel c) in~\cref{fig:circuit_implementation}.

We conclude with a remark on stray couplings, not captured by~\cref{eq:Hlab}, but most likely present in a superconducting circuit implementation. For instance, direct coupling~$g_{1-2}$  between the qubits cannot be suppressed by manipulating the coupler and, in that case,~$\mathrm{IPR}$ includes an additional term~$-2g_{1-2}^2/(\omega_{1}-\omega_{2})^2$ which bounds~$1-\mathrm{IPR}$. However, this bound can be conveniently lowered by detuning the qubits and improving circuit design such that~$g_{1-2}$ is minimized. It is also worth noticing that this interaction is typically very small compared to desired couplings. Another possibility for the presence of stray couplings is to have spurious qubit-NLR interactions. In this case, virtual two-qubit transitions mediated by the NLR are not exponentially suppressed. However, if the NLR is far detuned in frequency with respect to the qubits, these interactions can be greatly reduced. Finally, stray dispersive coupling between the bus and the NLR can also exist. However, this type of nonidealities are not particularly detrimental, as the resulting weak hybridization between the bus and the NLR does not prevent the suppression of two-qubit interactions. 

\end{document}